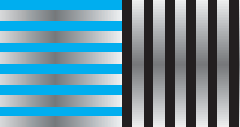

APPLIED RESEARCH

# Differentiable Cosimulator for Electrically Reconfigurable Structures

**JOHANNES MÜLLER, (Member, IEEE), DENNIS PHILIPP, AND MATTHIAS GÜNTHER**

Faculty of Physics, University of Bremen, 28359 Bremen, Germany
Fraunhofer MEVIS, 28359 Bremen, Germany

Corresponding author: Johannes Müller (jmller@uni-bremen.de)

This work was supported by U Bremen Research Alliance/AI Center for Health Care, and by the Federal State of Bremen.

**ABSTRACT** Electrically reconfigurable metasurfaces (MTS) for magnetic resonance imaging (MRI) can exhibit many degrees of freedom. Each tunable parameter affects the final response of the metasurface to an impinging magnetic field. Thus, the configuration of the electric parameters significantly affects the manner in which metasurfaces shape the overall magnetic field distribution. Owing to the high number of parameters and the mutual coupling, shaping the field in a target way can be a nontrivial and time-consuming challenge. This paper introduces a novel CUDA-enabled PyTorch-based framework (available open-source on github.com) designed for the gradient-based optimization of such reconfigurable electromagnetic structures with electrically tunable parameters. Traditional optimization techniques for these structures often rely on non-gradient-based methods, which limit their efficiency and flexibility. The proposed framework leverages automatic differentiation, facilitating the application of gradient-based optimization methods. This approach is particularly advantageous for embedding within deep learning frameworks, which enables sophisticated optimization strategies. We demonstrate the effectiveness of the framework through comprehensive simulations involving resonant structures with tunable parameters. The key contributions include the efficient solution to the inverse problem. The performance of the framework is validated using three different resonant structures: a single-loop copper wire (single unit cell) and an $8 \times 1$ as well as an $8 \times 8$ array of resonant unit cells with multiple inductively coupled unit cells (one-dimensional (1d) and two-dimensional (2d) metasurfaces). The results show precise control over an individual component of the magnetic field normal to the surface of each resonant structure, achieving target field strengths with minimal error.

**INDEX TERMS** Automatic differentiation, deep learning integration, differentiable simulation, electromagnetic structures, forward problem, gradient-based optimization, inverse problem, magnetic field control, MRI enhancement, PyTorch framework, reconfigurable metasurfaces, resonant structures, tunable parameters.

## I. INTRODUCTION

The application of properly configured resonant structures in magnetic resonance imaging (MRI) offers several advantages that significantly enhance imaging efficiency. Resonant electromagnetic structures, such as metasurfaces (MTS) and metamaterials (MTMs), manipulate the magnetic field in a well-defined manner, leading to improved performance metrics, such as an enhanced signal-to-noise ratio (SNR), increased imaging resolution, and reduced scanning time.

Resonant electromagnetic structures are static, tunable, or reconfigurable. Static structures maintain their properties over time. A significant enhancement in SNR [1], image resolution [1], and penetration depth [2], [3], [4] with preconfigured passive resonant structures is demonstrated. However, for specific use cases, the behavior of these structures may need to be adjusted to account for different loading conditions. Electrically tunable structures exhibit deliberately adjustable global behavior over time, allowing them to adapt to varying loading conditions [5] and adjust the modes to different frequencies [6]. They may also automatically detune to protect the patient from strong magnetic fields in transmit [7].

Reconfigurable resonant structures provide arbitrary spatiotemporal control of electromagnetic fields. The individual circuit parameters of these structures are adjusted to change their behavior. A local change in one cell can influence neighboring cells, making it challenging to accurately set the proper parameters for a target magnetic field distribution.

One of the first individually tunable resonant unit cells is presented in [8] highlighting the importance of finely-grained reconfigurability. Various tuning strategies, including electrical, optical, thermal, and mechanical adjustments, have been implemented for different applications [9], [10].

The potential of reconfigurable metasurfaces in MRI is demonstrated first by Wang et al. [11], where the enhancement ratio is selectively boosted by up to 28 times. A prototype with 14 degrees of freedom is proposed on other occasions [12], [13], [14].

### A. MOTIVATION

The deliberate control over the degrees-of-freedom allows for locally adjustable SNR-hotspots by properly tuning the capacitances of the resonant structure.

The capacitances $\mathbf{C}$ of reconfigurable electromagnetic structures and the resulting magnetic field $\mathbf{B}$ are related such that the mapping ($\mathbf{C} \rightarrow \mathbf{B}$) can be considered a forward problem. This relationship can be observed analytically, numerically and experimentally. The circuit model combined with the Biot-Savart law can be used if the structure is small compared to the wavelength. Analytical modeling is possible but is not agnostic to the geometry of the resonant structure. The inversion of the Biot-Savart law is challenging and can be inaccurate in many cases [15]. In some cases, it is posed as an optimization problem [16]. To the best of the authors knowledge, there is no analytical solution to the inverse of the Biot-Savart law for more complicated structures. Therefore, one model does not fit all resonant structures and cannot be reused for different structures.

The objective of this study is to exploit the full potential of reconfigurable resonant structures. This includes the inference of capacitance configurations of these structures that allow the realization of a target magnetic field distribution at given points in space.

### B. STATE-OF-THE-ART METHODS AND THEIR LIMITATIONS

Resonant structures, such as reconfigurable metasurfaces, exhibit many degrees of freedom that allow the targeted modification of the resulting magnetic field in MRI, and therefore, the adaption to a specific imaging goal and different loading conditions.

Simulations play a critical role in analyzing the interactions between electromagnetic fields, objects, resonant structures, and associated circuitry. The circuitry can significantly alter the resonant behavior of electromagnetic structures, thereby affecting the resulting electromagnetic fields. This interaction requires detailed analysis, which is why commercial software solutions, such as Dassault Systèmes CST Microwave Studio (CST MWS), include comprehensive simulation capabilities.

While (co-)simulators embedded in commercial software offer robust tools for circuit analysis, they often lack cost-efficient, standalone optimization of electrically tunable parameters. This limitation has led to the development and adoption of open-source solutions, particularly in MATLAB and Python environments. These solutions enable cost-effective, standalone calculations of electromagnetic fields concerning circuit parameters, providing a more flexible and accessible approach for researchers and engineers. A comparison of available solutions can be found in Tab. 1 and in Tab. 2.

A MATLAB-based solution [17] can be used with non-gradient-based optimization techniques, such as the Self-Organizing Migrating Algorithm (SOMA) or the fminsearch function. Similarly, an extensive Python-based circuit cosimulation framework was introduced [18]. This framework supports both 1-port and 2-port networks and leverages NumPy. However, a key limitation is the inability to perform gradient-based optimization because of the lack of automatic differentiation support in these packages. Other optimization techniques for resonant structures have been explored, as shown in [19]. Although this algorithm offers an efficient way to control the current phases in the structure, it is not applicable to certain structures. It also does not explicitly focus on the optimization of the magnetic or electric fields.

In comparison with existing solutions, the presented framework in this paper enables the exploitation of automatic differentiation. This allows fast and efficient calculation of gradients that directly point towards a lower error solution and, therefore, exploits gradient information in the optimization process. Opposed to algorithms normally used as e.g. genetic algorithms or SOMA, gradient descent does not require the evaluation of many candidate solutions in each epoch, which consumes less resources.

### C. NOVELTY

We propose a PyTorch-based framework that leverages automatic differentiation (AD) to facilitate the use of gradient-based optimization methods for tuning the circuit parameters within electromagnetic simulations. This approach is particularly compelling because of its potential for embedding within deep learning frameworks, thereby enabling sophisticated optimization strategies. The power of neural networks lies in their ability to learn from data during backpropagation, in which the gradients of the loss function are propagated backward through the network to update the weights. By employing a framework that inherently supports automatic differentiation, we can extend this powerful learning mechanism to electromagnetic simulations.

The proposed framework enables efficient, vectorized calculation of steady-state electric and magnetic fields in all three Cartesian field components, enabling direct control over the chosen fields while respecting the laws of physics.

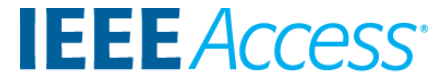

**TABLE 1.** Comparison of EM/circuit cosimulation tools for optimization of tunable parameters.

| | Commercial Solver (CST MWS) | CoSimPy [18] | Matlab Circuit cosimulation [17] |
|---|---|---|---|
| **License** | Proprietary | MIT | MIT |
| **Known limitations** | • Costly and closed-source<br>• No automatic differentiation<br>• No vectorization with respect to multiple imaging goals<br>• No direct integration into deep learning frameworks as e.g. PyTorch | • No out-of-the-box automatic differentiation<br>• No vectorization with respect to multiple imaging goals | • No out-of-the-box automatic differentiation<br>• No vectorization with respect to multiple imaging goals<br>• No possibility to attach different networks at excitation<br>• Limitation to capacitances only |
| **Supported Algorithms** | Trust Region Framework, Nelder Mead Simplex Algorithm, CMA Evolution Strategy, Genetic Algorithm, Particle Swarm Optimization, Interpolated Quasi Newton ; Available algorithms do not exploit automatic differentiation. | Compatible with Python algorithms that do not exploit automatic differentiation, such as genetic algorithms. | Compatible with MATLAB algorithms that do not exploit automatic differentiation, such as SOMA. |

**TABLE 2.** Qualitative comparison between non-gradient-based traditional optimization methods and the proposed PyTorch-based approach.

| | Traditional Optimization Methods (e.g., Genetic Algorithms, SOMA) | Proposed PyTorch-Based Approach |
|---|---|---|
| **Gradient availability** | No automatic differentiation; gradients are approximated or ignored | Full automatic differentiation via PyTorch/backpropagation |
| **Optimization** | Often heuristic/evolutionary search and therefore Non-gradient-based | Gradient-based |
| **Computational efficiency** | • Many candidate evaluations per iteration<br>• High runtime and resource usage | • Single backpropagation step per iteration<br>• Comparably lower runtime with less resource usage |
| **Scalability** | Number of possible parameters grows exponentially with dimensionality of the problem | Leverages gradient information to improve sample efficiency [20] |
| **Integration with deep learning** | No out-of-the box usage due to missing AD | Seamless embedding in PyTorch workflows (DL/MLOps) |
| **Convergence behavior** | • Stochastic; may require many evaluations<br>• No gradient direction to follow<br>• Therefore slower convergence compared to proposed approach | • Guided by exact gradients<br>• Faster and reliable convergence in practice (with suitable initialization) |

The computational effort required is linearly dependent on the number of data points and ports, which enhances scalability and performance [21]. In order to conduct the optimization, only a single full-wave simulation run is required. Furthermore, the framework is agnostic to both simulation software and geometry, eliminating the need for analytical modeling. Integration with PyTorch allows the use of PyTorch workflows and the incorporation of standard MLOps packages (e.g., mlflow), thereby facilitating streamlined and robust optimization processes. The user workflow is described in Sec. III and Fig. 2.

## II. PROBLEM STATEMENT

The proposed framework allows for the efficient solution of the forward problem $\mathbf{C} \rightarrow \mathbf{B}$. As the equation $\mathbf{B} = \mu\mathbf{H}$ holds and a phantom with permeability $\mu_r = 1$ is used, the following results are obtained for the $\mathbf{H}$-field. Gradient-descent-based optimization offers the possibility of efficiently tackling the inverse problem $\mathbf{H} \rightarrow \mathbf{C}$.

The central research question that shall be tackled using the presented framework can be formulated as follows: Is it possible to infer the circuit parameters C of multi dimensional resonant structures that cause a target magnetic field distribution. The circuit parameter C is a discretely tunable capacitance connected to the resonant structure. In this example the investigation will be limited the optimization of a target magnetic field distribution related to the configuration of tunable capacitances. The target magnetic field strength at each point in space is called $|\mathbf{H}|$ while the realized magnetic field strength is $|\hat{\mathbf{H}}|$.

The optimization goal is defined as follows:

$$\underset{\mathbf{C}\in\Omega}{\arg\min}\, \mathbf{MSE}\big(f(\mathbf{C}),\ \mathbf{H}\big) \tag{1}$$

$f(\mathbf{C})$ is the forward model of the physical system, $\mathbf{C}$ is the matrix of tunable parameters, and $\mathbf{H}$ is the target response. The optimization objective is to find a parameter configuration within a feasible domain $\Omega$ that minimizes the mean-squared-error (MSE). The feasible domain in this investigation is not bound.

The problem under investigation is an inverse one. Inverse problems can be well- or ill-posed. Hadamard defined a well-posed problem as a problem that has to fulfill the following widely known criteria: 1) existing solution; 2) unique solution; and 3) continuous dependence on input data. A problem is ill-posed if one or more of the criteria are violated.

## III. METHODS

Full-wave electromagnetic (EM) multiport simulations accurately model the behavior of complex electromagnetic systems. The simulation process involves the individual excitation of the system through multiple ports, each treated as an impedance element — a current source with an inner impedance capable of both exciting and absorbing power [22]. During the simulation, the current source becomes active, stimulating each port individually. S-matrix and EM fields of the systems response are exported, providing insight into the interaction of individual inputs with respect to the outputs of the system.

The resulting data encompasses the multi-port S-parameter matrix and three-dimensional (3D) electromagnetic fields, representing the behavior of the system. The S-parameters and EM fields can then be used for subsequent analysis, leveraging the capabilities of tools such as CoSimPy [18] or the framework presented in this study. The calculations in this framework are based on the method proposed in [18] and therefore follow similar definitions where appropriate. In the following, the S-matrix extracted from the simulation software that contains the scattering parameters of the unconnected resonant structure shall be named $\mathbf{S}_0$ the exported fields can be either **B** or **H** and **D** or **E**.

Modelling of the circuitry is also required. In this study, RC circuits were used. Other types of circuits can also be used. A simple RC-circuit can be modelled by

$$\mathbf{Z}_C = \mathbf{R} - j\frac{1}{\omega \mathbf{C}}, \tag{2}$$

which relates to the S-parameters **S** using

$$\mathbf{S} = \frac{\mathbf{Z_C} - \mathbf{Z_0}}{\mathbf{Z_C} + \mathbf{Z_0}}, \tag{3}$$

with $\mathbf{S}, \mathbf{Z_C}, \mathbf{C} \in \mathbb{C}^{n_{batch} \times n_f \times n_0}$.

$\mathbf{Z_0}$ is the characteristic impedance of an individual port. For simplicity, assume $n_f = n_{batch} = 1$ and omit these dimensions in the following discussion.

The circuitry must be connected to the modelled resonant structure. Therefore a new tensor $\mathbf{S_C} \in \mathbb{C}^{(n_0+n_1)\times(n_0+n_1)}$ is built that can be separated into four parts [18]. See Fig. 1.

- $\mathbf{S_{C_{11}}} \in \mathbb{C}^{n_0 \times n_0}$ represents the reflection from the output perspective of the networks that connect to the resonant structure.
- $\mathbf{S_{C_{22}}} \in \mathbb{C}^{n_1 \times n_1}$ represents the reflection from the input perspective of the networks that connect to the resonant structure.
- $\mathbf{S_{C_{12}}} \in \mathbb{C}^{n_0 \times n_1}$, $\mathbf{S_{C_{21}}} \in \mathbb{C}^{n_1 \times n_0}$ represent the transmission from input to output and vice-versa.

$n_0$ is being assigned to the number of ports of the resonant structure ($\mathbf{S_0}$) and $n_1$ is being assigned to the number of external ports ($\mathbf{S_C}$), respectively.

As a simplified example, for 1-port networks, only values on the diagonal of $\mathbf{S_{C_{11}}}$ are assigned as we connect the output side to the resonant structure. These can only reflect power to a certain degree on this one port and do not transmit energy to the input ports of $\mathbf{S_C}$.

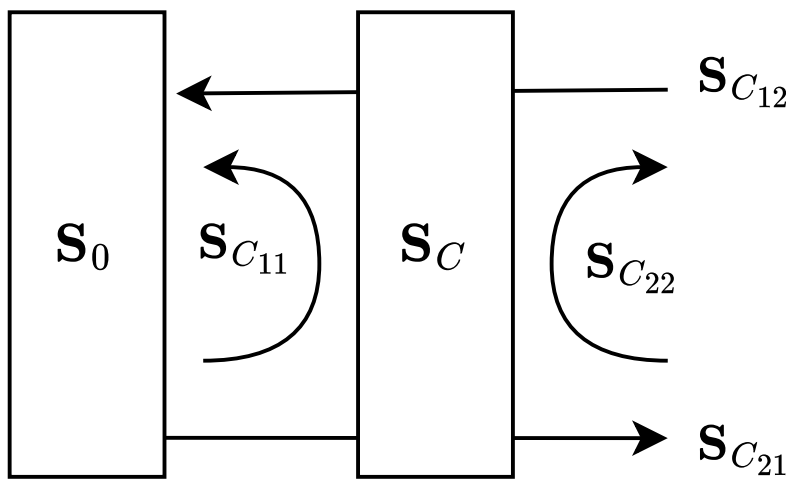

**FIGURE 1.** Visualization of the connection of the networks contained in $\mathbf{S_C}$ to the resonant structure $\mathbf{S_0}$.

To calculate the resulting fields, it is critical to consider the incident voltage wave at each port.

The incident voltage wave at each port of the passive resonant structure $\mathbf{V}^+_{\mathbf{res}} \in \mathbb{C}^{n_0 \times n_1}$ can be described as

$$\mathbf{V}^+_{\mathbf{res}} = \mathbf{S}_{C_{11}}\mathbf{V}^-_{\mathbf{res}} + \mathbf{S}_{C_{12}}\mathbf{V}^-_{\mathbf{exc}} \tag{4}$$

with $\mathbf{V}^-_{\mathbf{exc}} = \sqrt{\mathbf{Z}'_{\mathbf{exc}}\mathbf{P}}$ being the outgoing voltage wave from the excitation. $\mathbf{Z}'_{\mathbf{exc}}$ is the characteristic impedance at excitation. $\mathbf{P}$ is often assumed to be 1 W while $\mathbf{Z}'_{\mathbf{exc}}$ is often chosen to be 50 Ω for each port. These assumptions will also be made for the results generated in the following.

$\mathbf{V}^+_{\mathbf{res}}$ is the incident voltage wave in the resonant structure and $\mathbf{V}^-_{\mathbf{res}}$ is the reflected voltage wave from the passive structure. The relation

$$\mathbf{V}^-_{\mathbf{res}} = \mathbf{S}_0\mathbf{V}^+_{\mathbf{res}} \tag{5}$$

holds. This results in

$$\frac{\mathbf{V}^+_{\mathbf{res}}}{\mathbf{V}^-_{\mathbf{exc}}} = (\mathbf{I} - \mathbf{S}_{C_{11}}\mathbf{S}_0)^{-1}\mathbf{S}_{C_{12}}. \tag{6}$$

with **I** being the identity matrix. This description is consistent with [18].

The resulting voltage-wave-ratio from Eq. 6 can then be used to linearly superimpose the individual electric or magnetic fields with respect to their amplitude and phase to one common electric or magnetic field. For example, the B-field can be described as follows:

$$\mathbf{B_{combined}} = \mathbf{B}\frac{\mathbf{V}^+_{\mathbf{res}}}{\mathbf{V}^-_{\mathbf{exc}}} \tag{7}$$

With the Field $\mathbf{B} \in \mathbb{C}^{n_{points} \times 3 \times n_0}$ and $\mathbf{B_{combined}}$ being a $(n_{points}, 3, n_1)$ tensor. This is comparable to the results of [18] and [21].

Gradients of the magnetic field $\mathbf{B_{combined}}$ with respect to the tunable parameters **C** can be expressed as follows:

$$\frac{\partial \mathbf{B_{combined}}}{\partial \mathbf{C}} = \mathbf{B}\frac{(\mathbf{V}^+_{\mathbf{res}}/\mathbf{V}^-_{\mathbf{exc}})}{\partial \mathbf{C}} = \mathbf{B}\frac{\mathbf{A}^{-1}\mathbf{S}_{\mathbf{C}_{12}}}{\partial \mathbf{C}} \tag{8}$$

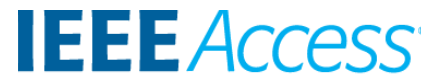

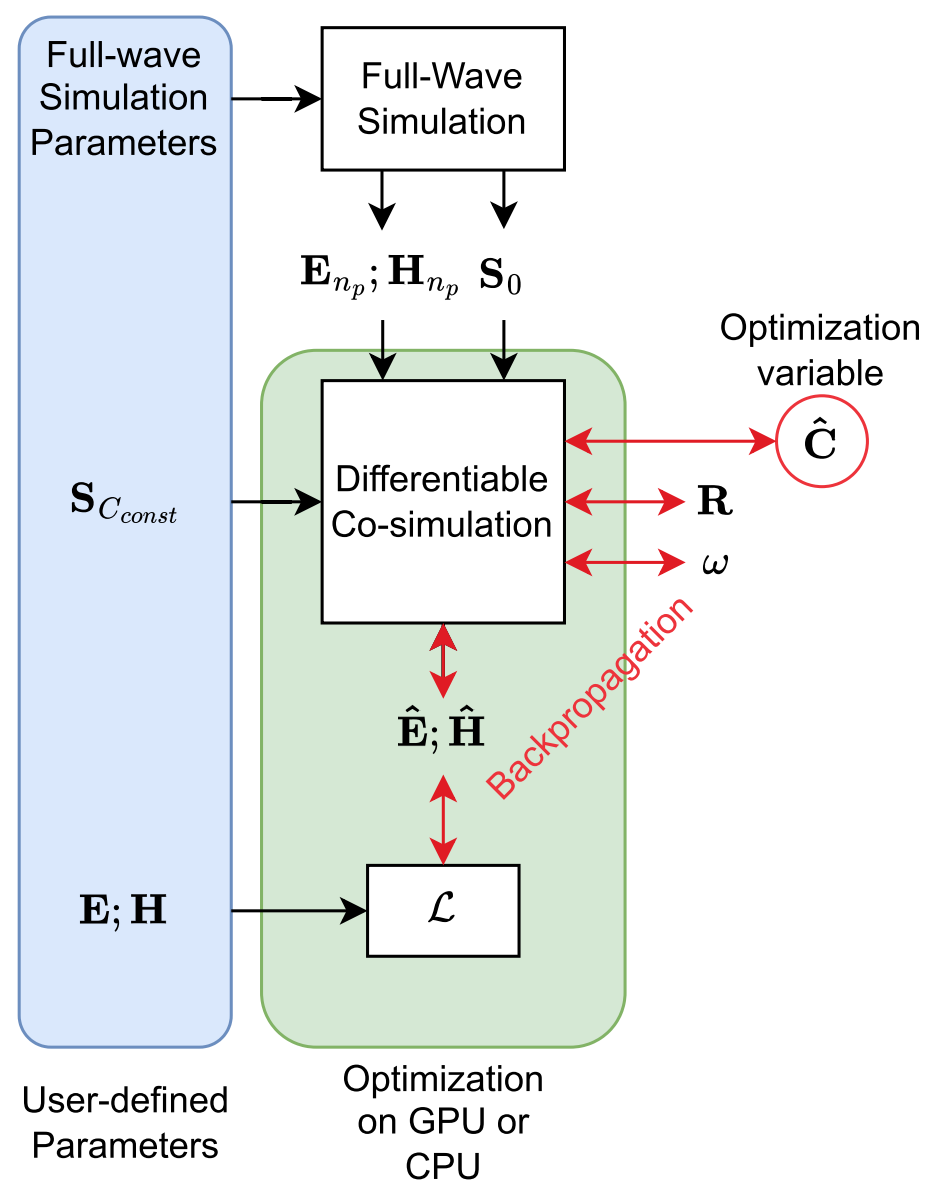

**FIGURE 2.** S-parameters $S_0$ and field-data $H_{np}$ can be exported from a full-wave simulation and fed into the differentiable cosimulation framework. All input and output data are torch tensors that allow gradient tracking and CUDA-enabled GPU processing. The user chooses the full-wave simulation parameters, multiport network scattering parameters $S_{C_{const}}$ and the target magnetic or electric field distribution $E; H$. Backpropagation enables gradient descent-based optimization and allows for the efficient adjustment of optimization variables **R**, **L** (not shown here), **C** or $\omega$ to minimize the loss function $\mathcal{L}$. The framework therefore facilitates finding lower-error solutions to an inverse problem.

with $\mathbf{A} = \mathbf{I} - \mathbf{S}_{\mathbf{C}_{11}}\mathbf{S}_0$. Considering the basic identity from [23] with reference to [24], this results in

$$= \mathbf{B}(-\mathbf{A}^{-1}\frac{\partial \mathbf{A}}{\partial \mathbf{C}}\mathbf{A}^{-1}\mathbf{S}_{\mathbf{C}_{12}} + \mathbf{A}^{-1}\frac{\partial \mathbf{S}_{\mathbf{C}_{12}}}{\partial \mathbf{C}}) \tag{9}$$

and finally leads to

$$= \mathbf{B}\mathbf{A}^{-1}(\frac{\partial \mathbf{S}_{\mathbf{C}_{11}}}{\partial \mathbf{C}}\mathbf{S}_0\mathbf{A}^{-1}\mathbf{S}_{\mathbf{C}_{12}} + \frac{\partial \mathbf{S}_{\mathbf{C}_{12}}}{\partial \mathbf{C}}). \tag{10}$$

$\mathbf{B}_{\mathbf{combined}}$ is related to the signal enhancement $\beta$ by normalization to a reference that does not relate to the tunable parameters, as described in Sec. IV-B. The tunable parameters $\mathbf{C}$ are related to the S-parameters by Eq. 3.

### A. OPTIMIZATION PIPELINE

The general user workflow is a two-step process: i) full-wave simulation; ii) differentiable cosimulation with backpropagation.

The differentiable simulator requires s-parameter and field data exported from a full-wave simulation. Data processed by the pipeline can be pushed to the graphical processing unit (GPU) beforehand. All required calculations can be conducted on the GPU, facilitating application in deep learning. Backpropagation then allows for efficient gradient-based optimization. Fig. 2 visualizes and describes the interfaces of the differentiable simulator used in this study.

The variable $\mathbf{S}_\mathbf{C}$ represents an intermediate multiport network (comparable to [18]) that manages the connection from the resonant structure ($\mathbf{S}_\mathbf{0}$) with its $n_0$ ports to the output ports of the structure connected to circuitry ($n_1$). Therefore, the scattering tensor $\mathbf{S}_\mathbf{C}$ links the resonant structure to the networks.

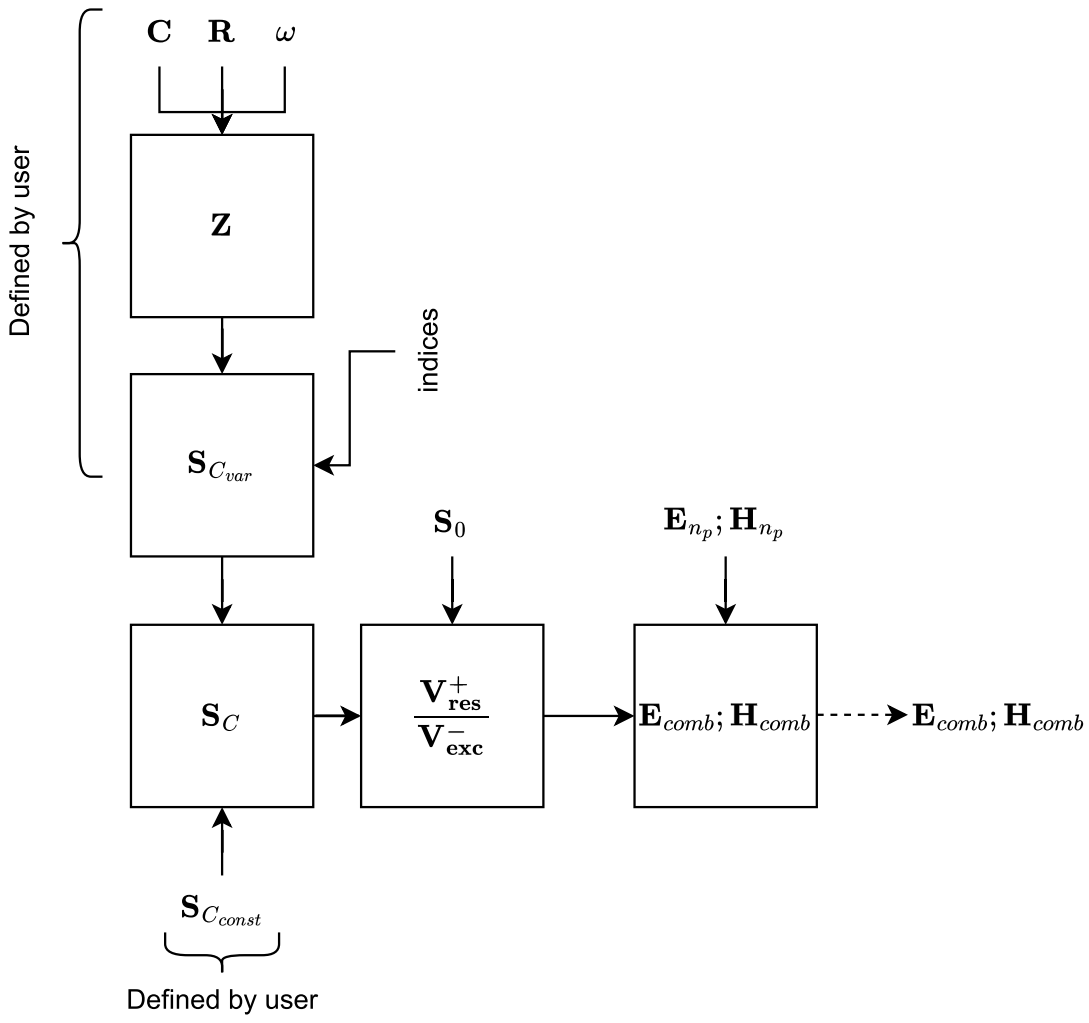

**FIGURE 3.** The fundamental pipeline calculates fields from given variables $\mathbf{S_C}$ and $\mathbf{S_0}$ as well as the fields that shall be combined $\mathbf{E}_{np}$, $\mathbf{H}_{np}$. The user can freely choose how to pass the individual elements of $\mathbf{S_C}$ to the fundamental pipeline. In this case it might be useful to map specific circuit parameters to individual elements of $\mathbf{S_C}$. Therefore a preprocessing procedure is written and used in combination with the fundamental calculation pipeline.

The optimization pipeline allows for the optimization of any arbitrary S-parameter within the $\mathbf{S}_\mathbf{C}$ tensor. Therefore, the pipeline is designed to offer maximum flexibility in defining custom preprocessing steps. The internal workflow of the pipeline has been visualized in Fig. 3.

1) **User-Defined Preprocessing**: User-defined preprocessing of data that results in S-parameters. The user can build any preprocessing step that relates to the individual elements of $\mathbf{S}_{C_{var}}$. The custom preprocessing step must output a three-dimensional tensor $(n_{batch}, n_f, n_{S-params})$. The user defines the *indices* that map each element of this output tensor to the corresponding positions in the $\mathbf{S}_{C_{var}}$ tensor. *indices* contains two tensors with indices for $x$ and $y$ position assignment in $\mathbf{S}_{C_{var}}$. The first tensor contains the indices for the rows ($x$), and the second contains the indices for the columns ($y$). This ensures that the zero-valued $\mathbf{S}_{C_{var}}$ tensor is populated with relevant values, as determined by the user-defined indices.
2) **Constant S-parameter Definition**: Users also define an $\mathbf{S}_{C_{const}}$ tensor containing constant S-parameters that remain unchanged during the optimization. This allows portions of the multiport network to remain fixed, while other parts (defined by $\mathbf{S}_{C_{var}}$) are optimized.
3) **Combination of $\mathbf{S}_{C_{var}}$ and $\mathbf{S}_{C_{const}}$**: $\mathbf{S}_\mathbf{C} = \mathbf{S}_{C_{var}} + \mathbf{S}_{C_{const}}$ is an element-wise linear superposition of S-matrices describing constant and tunable subcircuits.

This creates the complete $\mathbf{S_C}$ tensor that defines the behavior of the multiport network in the simulation.

4) **Field Combination and Simulation**: After constructing the $\mathbf{S_C}$ tensor, the pipeline proceeds with the field calculations, ultimately combining the electromagnetic fields as part of the final simulation output.

### *B. NUMERICAL SETUP*

The structure can be modelled using a full-wave electromagnetic solver toolkit, such as Sim4Life [25] or CST MWS.

In the following example, three different resonant structures were constructed and simulated in a multiport simulation utilizing the Frequency Domain solver in CST MWS 2024. All structures were based on coupled split-ring geometries with different unit cell sizes or different numbers of cells. To demonstrate the versatility of the framework, they were excited in two different ways: an untuned 1-port excitation coil and a tuned 2-port birdcage coil. The chosen simulation frequency is the Larmor frequency of proton MRI at 2.9 Tesla, which corresponds to approximately 123.5 MHz. All the resonant cells have a wire width of 1 mm on the FR-4 substrate. The cells rest on a phantom ($\varepsilon = 70$; $\sigma = 0.7\,S/m$) that introduces loading. The magnetic field component extracted is always the component normal to the surface of the resonant structure. The chosen boundary conditions for all shown simulations are ''open (add space)''.

The first resonant structure consists of a single loop copper wire with an inner radius of 30 mm that is split by one port. A second port is connected to a perfect-electric-conductor (PEC) loop coil with an outer radius of 45 mm and a wire diameter of 1 mm that shall drive the resonant structure during cosimulation with 1 W. The first port is connected to a tunable RC-series network. The setup is illustrated in Fig. 4.

The second resonant structure contains a 8 × 1 line of coupled unit cells with an outer diameter of 28 mm and a center-to-center distance of 30 mm. It is exposed to a circularly polarized magnetic field generated by a tuned 2-port birdcage coil connected to two discrete ports. The two discrete ports were powered with 1 W and a phase shift of 90° to one another. The setup is illustrated in Fig. 5.

For the third structure, the 1d-line is extended to 8 × 8 allowing more degrees of freedom in shaping the field in space. While the second structure is excited by a tuned birdcage coil, the 8 × 8 structure is excited by an excitation coil with a diameter of 300 mm located at a distance of 150 mm above the substrate. The complete setup is shown in Fig. 6.

### *C. OPTIMIZATION CONFIGURATION*

For the first resonant structure, the magnetic field is measured at a distance of $z = -30$ mm. For the other resonant structures, the points of interest are located precisely at the center of each unit cell, offset by 0.8 times the cell diameter within the phantom. The points of interest are also called collocation points when discussing the optimization problem.

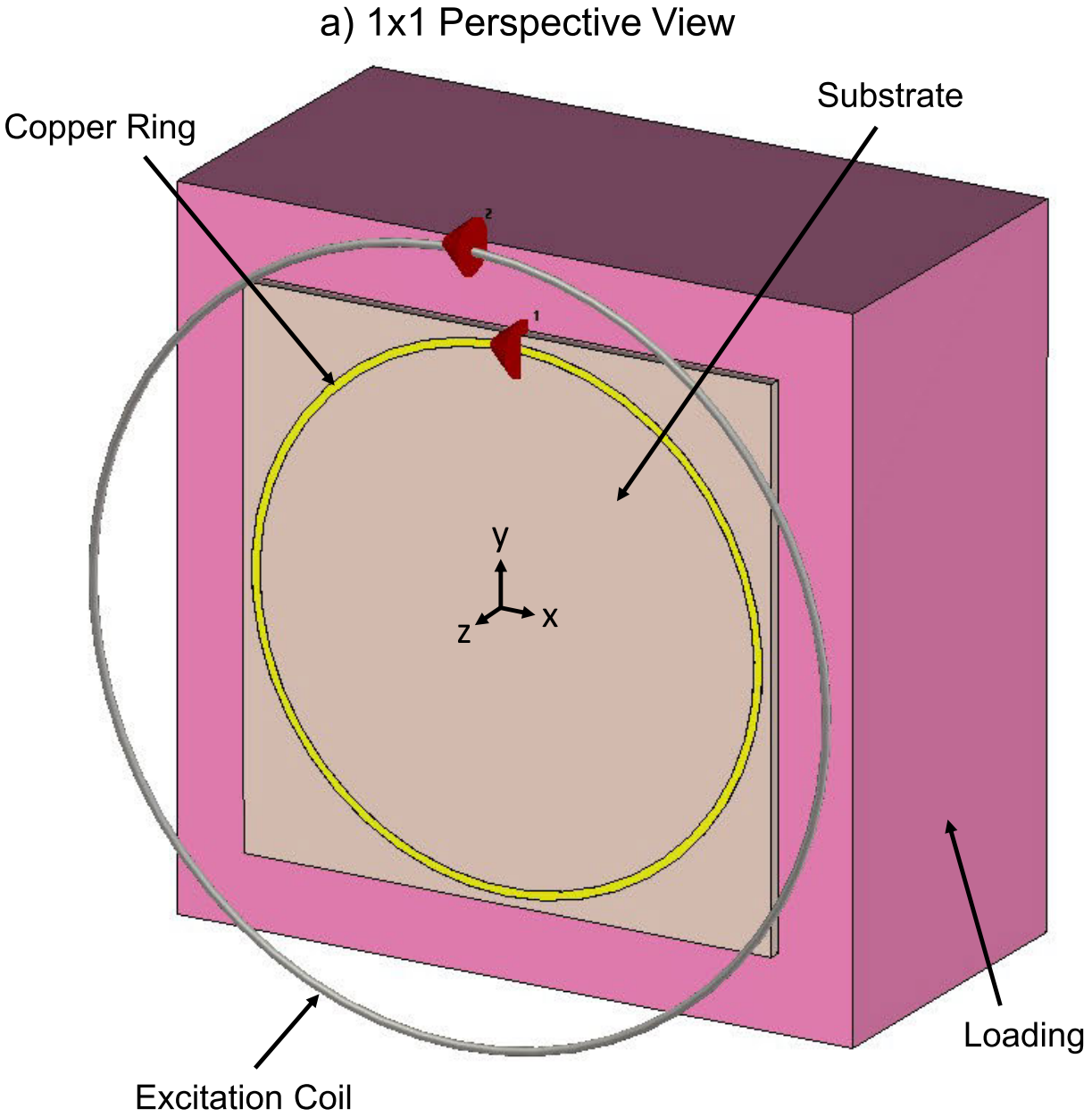

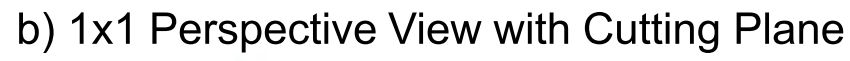

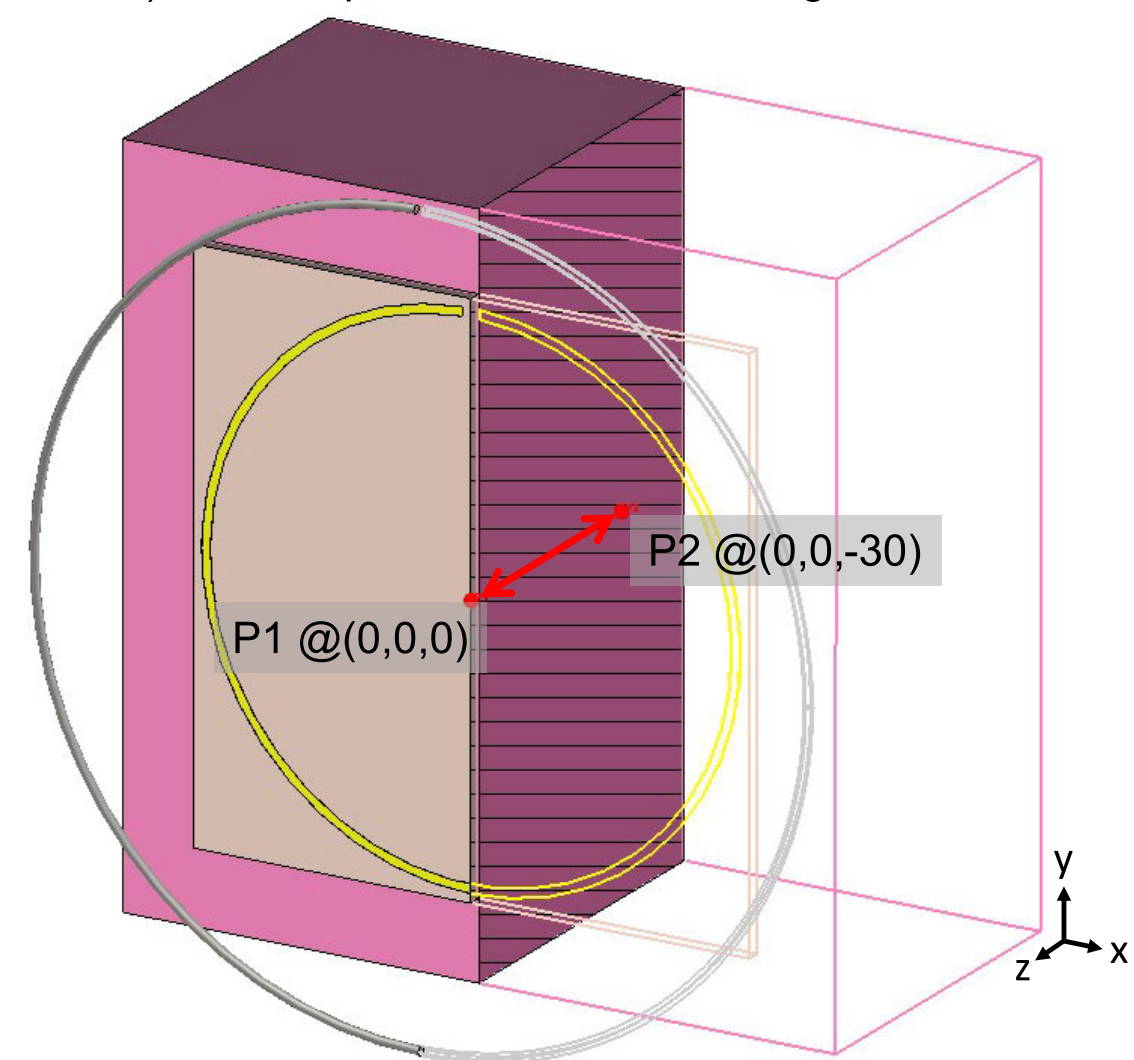

**FIGURE 4. 0D single-cell resonant structure with one tuning network is shown in a) and b). The structure rests on a phantom. Port 1 is connected to a tunable RC network. The PEC excitation loop coil was driven by port 2. The coil excited the resonant structure during the cosimulation. The origin of the coordinate system is at P1 in b). The point of observation is at P2 in b).**

During the gradient descent-based optimization, the Adaptive Moment Estimation (ADAM) optimizer is utilized with a maximum of 2000 epochs and a learning rate of $1 \times 10^{-13}$. Since typical capacitance values are on the order of $1 \times 10^{-12}$, the learning rate is chosen to be at least one order of magnitude smaller to ensure that updates remain comparable to realistic changes in the capacitance configuration during learning. The Adam optimizer, a standard method commonly used in deep learning, offers features such as an adaptive learning rate and momentum [26]. The performance was evaluated using the mean squared error (MSE). The optimization was conducted using a central-processing-unit (CPU)-only to ensure exact reproducibility. The CPU used for the optimization was an Intel Core i7 13700KF with 64GB RAM. GPU-based acceleration may also be used.

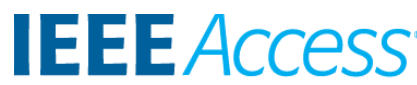

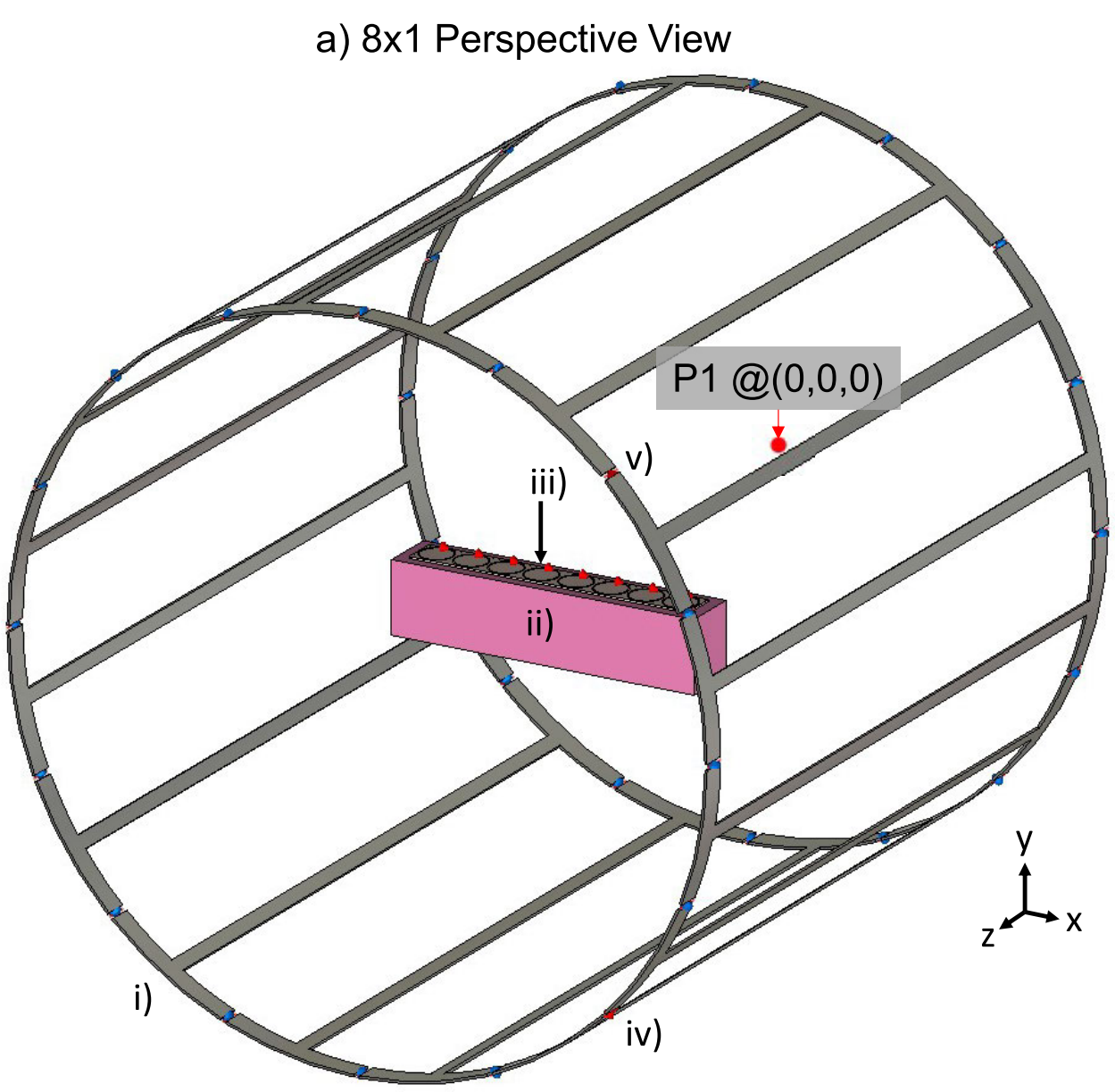

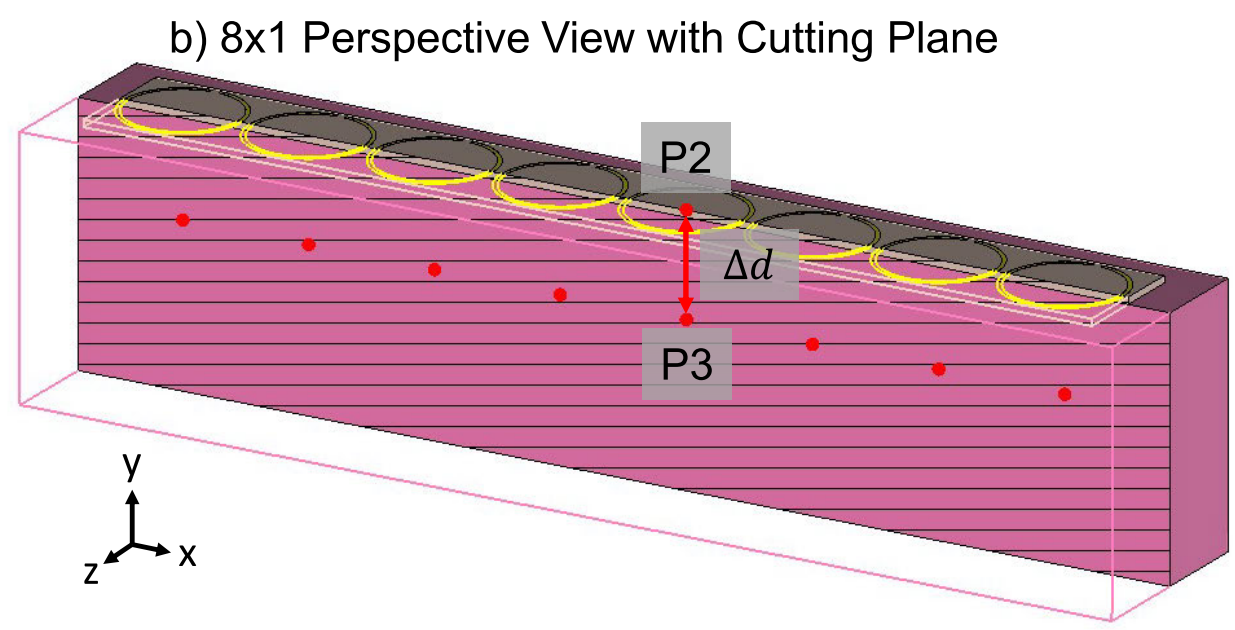

FIGURE 5. a) 1D resonant structure exposed to a circularly polarized magnetic field of a birdcage coil i). Ports 1-2 (v)-iv)) are connected to the birdcage coil and driven with a 90° phase shift to one another. The remaining eight ports of resonant structure iii) are connected to 1-port RC-Networks. The resonant structure was loaded ii). The radio frequency shielding (RF-shielding) is hidden to allow the reader to obtain a complete view of the setup. Point P1 represents the origin of the coordinate system. b) shows a cutting plane view of the 8 × 1 array. The observation points have distance Δd from the center of each unit cell. The points of observation also apply to an 8 × 8. Point P2 indicates the center of one of the cells. Point P3 is an exemplary observation point.

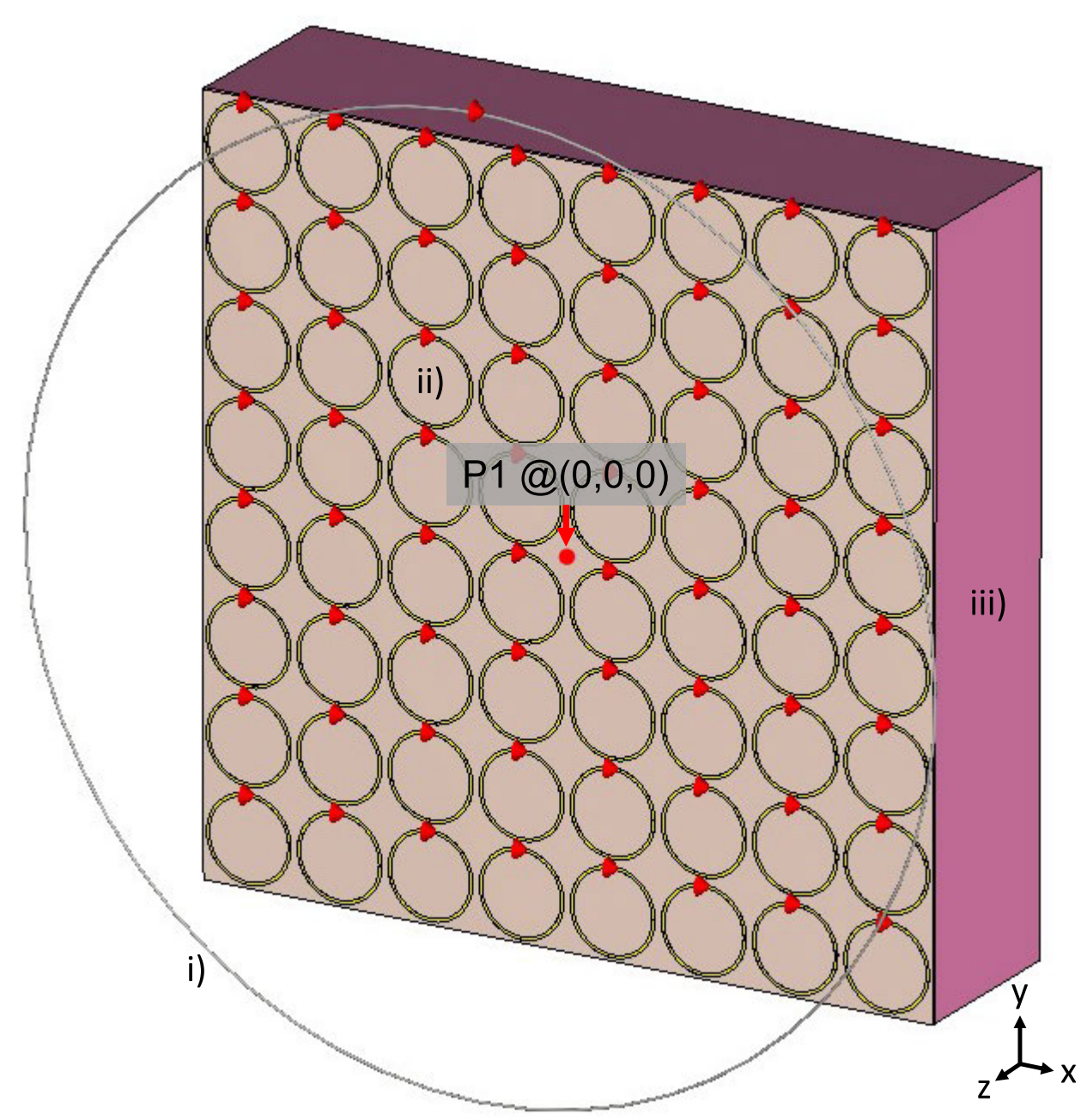

FIGURE 6. 2D resonant structure consisting of an array of 8 × 8 unit-cells. The structure rests on iii) a phantom. The phantom and resonant structures were exposed to the magnetic field of i) an excitation coil. Port 1 was connected to the excitation coil. The remaining 64 ports are connected to 1-port RC-Networks. The placement of the observation points are comparable to Fig. 5 b) with a distance of z=-22.4 mm from the isocenter of each cell in the phantom.

## IV. RESULTS

### A. RECONFIGURABLE 0D UNIT-CELL OPTIMIZATION

The system model of a simple capacitively loaded Ring (CLR) without excitation can be described as follows:

$$Z = R + j\omega L - j\frac{1}{\omega C} \tag{11}$$

In resonance, $Im\{Z\} = 0$ holds. The inductance $L$ and potential mutual coupling $M$ can be extracted from the simulation. For n-dimensional cases, the following relationship holds:

$$L + M = -\frac{1}{\omega \sum_{i=0}^{n_{cell}} \Im(Y_i)}. \tag{12}$$

Therefore, one can solve for $C_r$ as

$$C_r = \frac{1}{\omega^2(L + M)}. \tag{13}$$

For this specific single-cell case, M can be neglected. Using Eq. 13 a capacitance of 8.39pF is extracted. To use plausible values for $|\mathbf{H}_z|$, the fields for capacitances of 0.1 pF and 30 pF were extracted. This results in a magnetic field strength of approximately 0.2 A/m up to 2.3 A/m.

The resulting error is $8.03 \times 10^{-18}$ for epoch=931. In tendency, a longer optimization time results in better convergence. Owing to the nature of the Adam optimizer, the end result is still subject to fluctuation.

### B. RECONFIGURABLE 1D MTS OPTIMIZATION

In the following, $|H_y|_{MTS}$ (simulation with resonant structure) is optimized relative to $|H_y|_{ref}$ (simulation without a resonant structure). Therefore, optimization was conducted directly on the magnetic field enhancement factor to demonstrate the broader applicability of the algorithm at different scales. The enhancement factor shall be defined as $\beta = |H_y|_{MTS}/|H_y|_{ref}$

A capacitance configuration shall be found to demonstrate the following cases: 1) peak; 2) trough; 3) Hadamard patterns; 4) homogenization; and 5) arbitrary field shaping.

A total of 1000 exemplary samples of arbitrary patterns and 32 systematic patterns were generated. The mapping $\mathbf{C} \mapsto \beta$ is of the form $(f : \mathbb{R}^8 \mapsto \mathbb{R}^8)$. The target $\beta$ ranged from 1 to the mean of the magnetic field enhancements of all cells at resonance (=2.18).

The successful realization of a representative selection of 1032 target field enhancements $\beta$ is demonstrated in Fig. 7.

For all $\beta$, the algorithm identified an appropriate capacitance configuration that resulted in a low error between the target $\beta$ and the realized $\hat{\beta}$ (s. Fig. 7 a-d) + f)). All realizations were achieved with a low error of $\leq 4.4 \times 10^{-5}$. The error distribution is illustrated in Fig. 7 f) The convergence plot in Fig. 9 a) exhibits convergence to lower errors.

As shown in Fig. 8, across all samples, a tendency towards higher errors can be observed when aiming for a higher $\hat{\beta}$ or highly spatially alternating (high-frequency) patterns.

### C. RECONFIGURABLE 2D MTS OPTIMIZATION

To prove the general usability of this approach, the field distribution $|\mathbf{H}_z|$ of an $8 \times 8$ array of resonant unit cells is optimized in the following. The mapping $\mathbf{C} \mapsto |\mathbf{H}_z|$ is of the form $f : \mathbb{R}^{64} \mapsto \mathbb{R}^{64}$. As shown in Eq. 12 and Eq. 13 the initial parameters for the optimization can be extracted analytically. The maximum $|\mathbf{H}_z|$ at resonance cannot be realized consistently over the entire field distribution. To guarantee the realizability of the magnetic field distribution, only approximately half of the maximum of a single resonant cell was taken as the upper boundary. The other boundaries were calculated analogously to the $8 \times 1$ case.

Four different systematic patterns were tested for the $8 \times 8$ prototype: 1) capital letters [A-Z]; 2) digits [0-9]; 3) homogenization; and 4) Hadamard matrix of order eight. This resulted in 45 individual samples.

The optimization converges to an error of $8.37 \times 10^{-8}$ in epoch 1996. Overall, the optimization procedure took 53 seconds wall-clock time.

Fig. 10 illustrates the electromagnetic field realizations corresponding to the letters "M-E-V-I-S". The combined field magnitudes $|\hat{H}_z|$, as shown in Fig. 10 a), have been computed and verified using CST MWS 2024. The intended letter shapes were successfully generated at their respective spatial coordinates.

Although the minimum and maximum field intensities at the target locations are consistently achieved, some regions between the specific collocation points exhibit field strengths that deviate from the minimum and maximum values given during the optimization process, as they are not explicitly considered.

However, within the region of interest (−105 mm to 105 mm), the letters are clearly distinguishable. The optimizer converges straight towards a good solution, as exemplarily demonstrated for the letter "S" in Fig. 11 for the magnetic field distribution controlled by the capacitance configurations shown in Fig. 12. The shape of the letter is already visible after a few epochs. The convergence plot shown in Fig. 9 b) exhibits stable convergence towards lower errors. These results underscore the capability of the optimization procedure in finding low-error solutions for the formation of target magnetic field patterns.

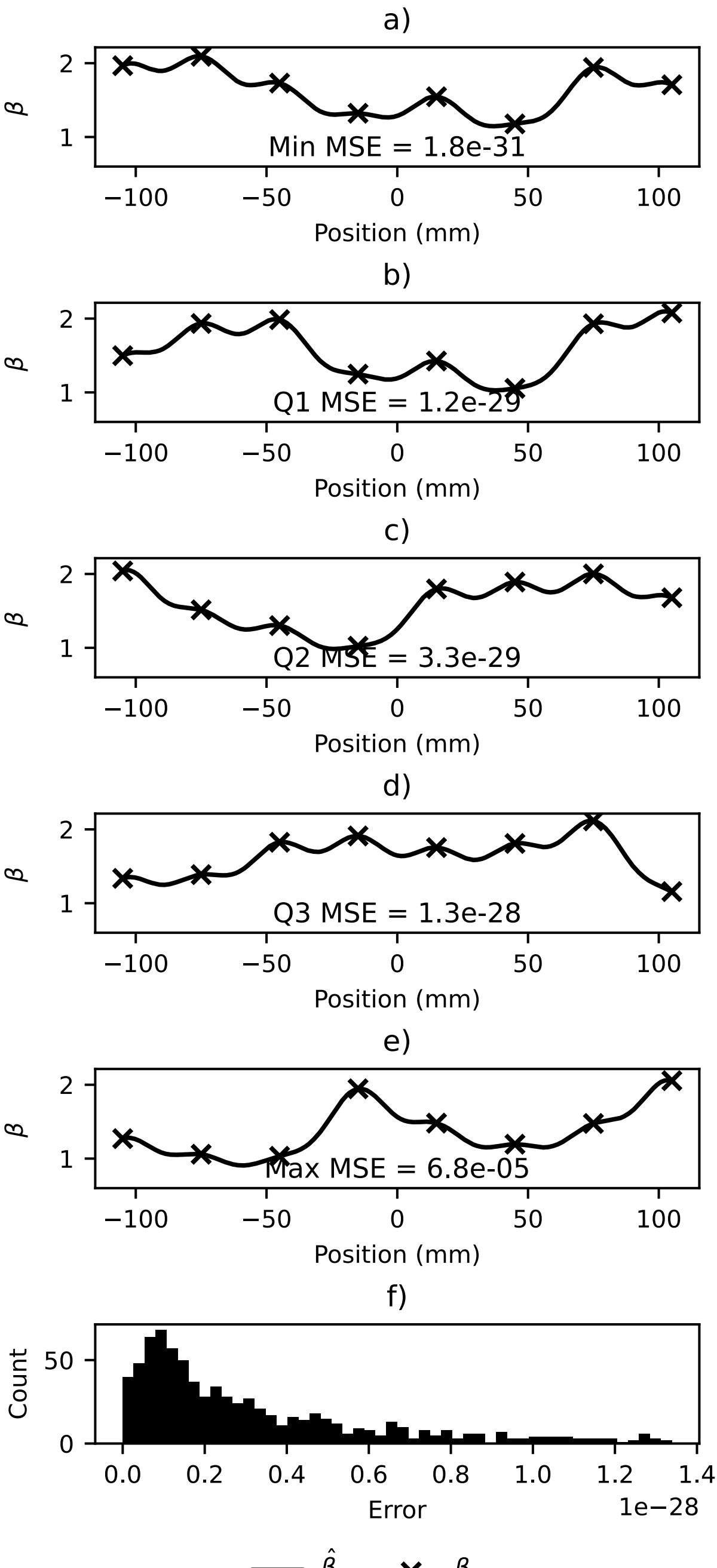

**FIGURE 7. Distribution plot of the magnetic field enhancement for the 8 × 1 case. Individual samples with the lowest a), first quartile b), median c), third quartile d), and highest error e) are shown. The target magnetic field enhancement $\beta$ is compared and quantified to the realized magnetic field enhancement $\hat{\beta}$ using the MSE at the collocation points. In f) a distribution plot shows the error distribution for all 1032 samples in the range from zero to the third quartile. There is at least one outlier with an error of $4.4 \times 10^{-5}$.**

The fields realized in CST MWS 2024 agreed with the target magnitude at the specified collocation points

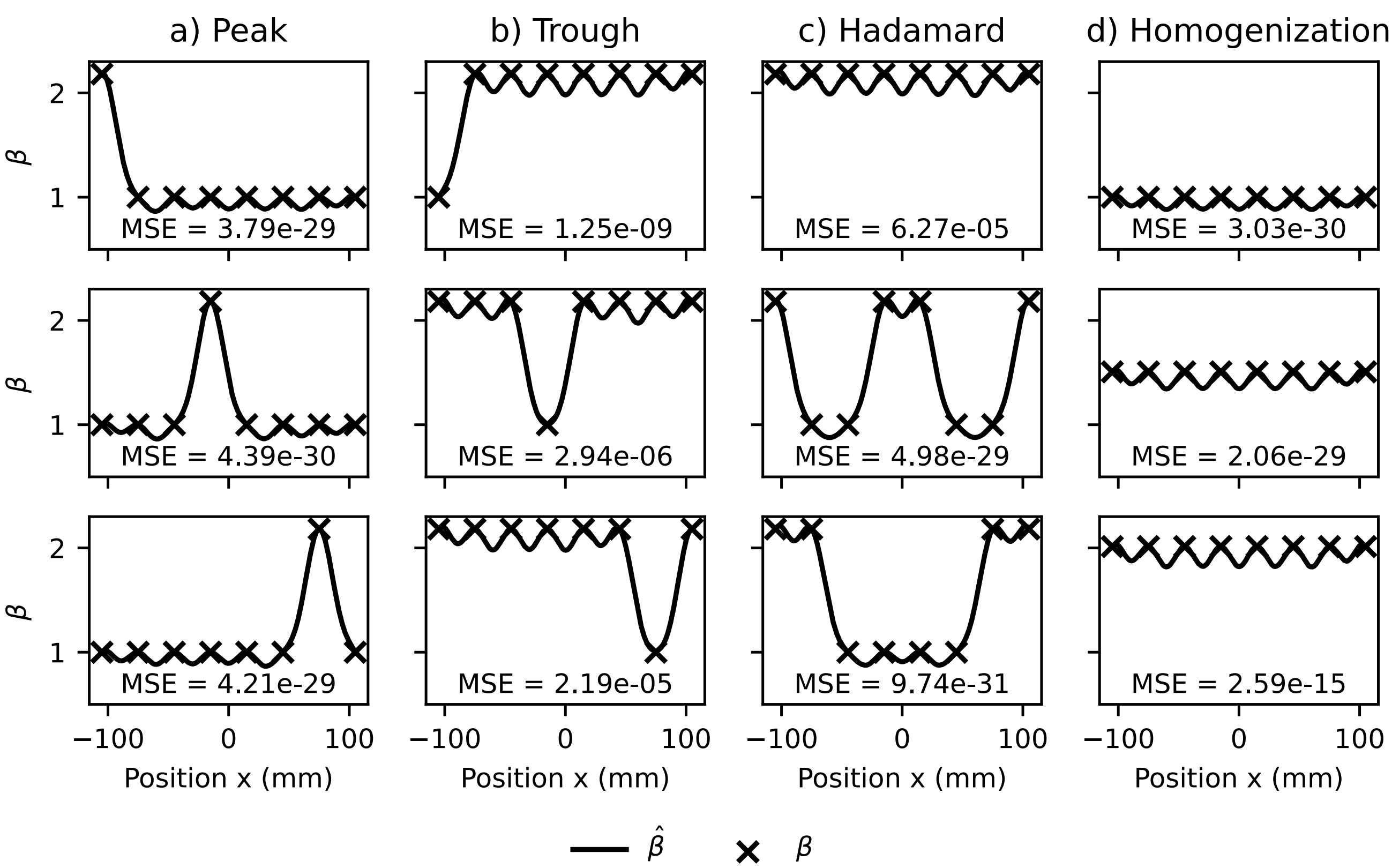

**FIGURE 8.** Demonstration of different systematic target and optimized magnetic field enhancement patterns $\hat{\beta}$ with a low error of less than $4.4 \times 10^{-5}$ to $\beta$. In a), a shifting peak moving from the leftmost center of the cell to the rightmost center of the cell is shown. In b) the $\beta$ is inverted. A shifting trough is shown in the figure. In c) Hadamard patterns are realized. In d) different strengths of homogenization are shown.

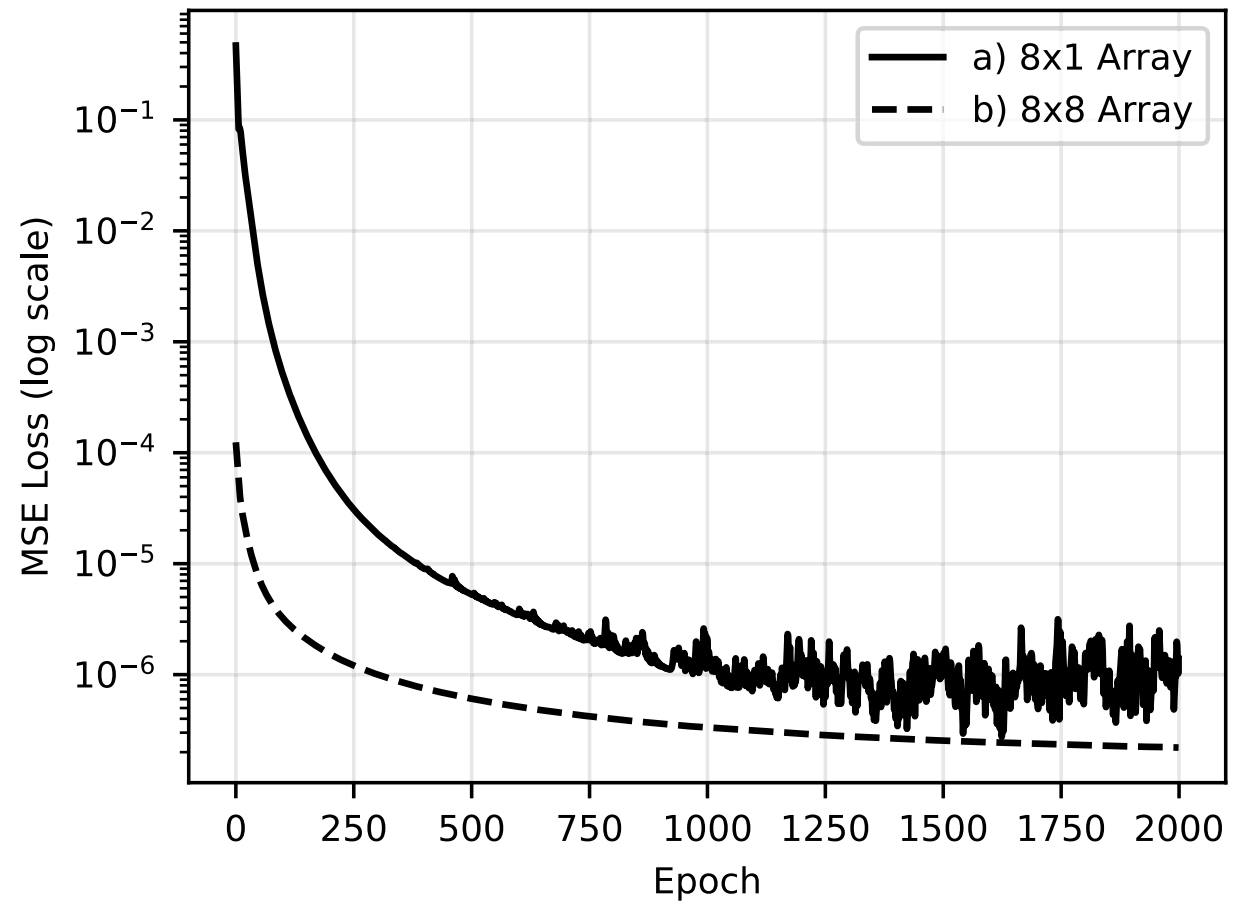

**FIGURE 9.** Log-scaled convergence plot of the gradient descent approach optimizing the tunable parameter on an a) 8 × 1 array with 1032 samples and b) an 8 × 8 array with 45 samples. Both plots exhibit convergence towards lower errors.

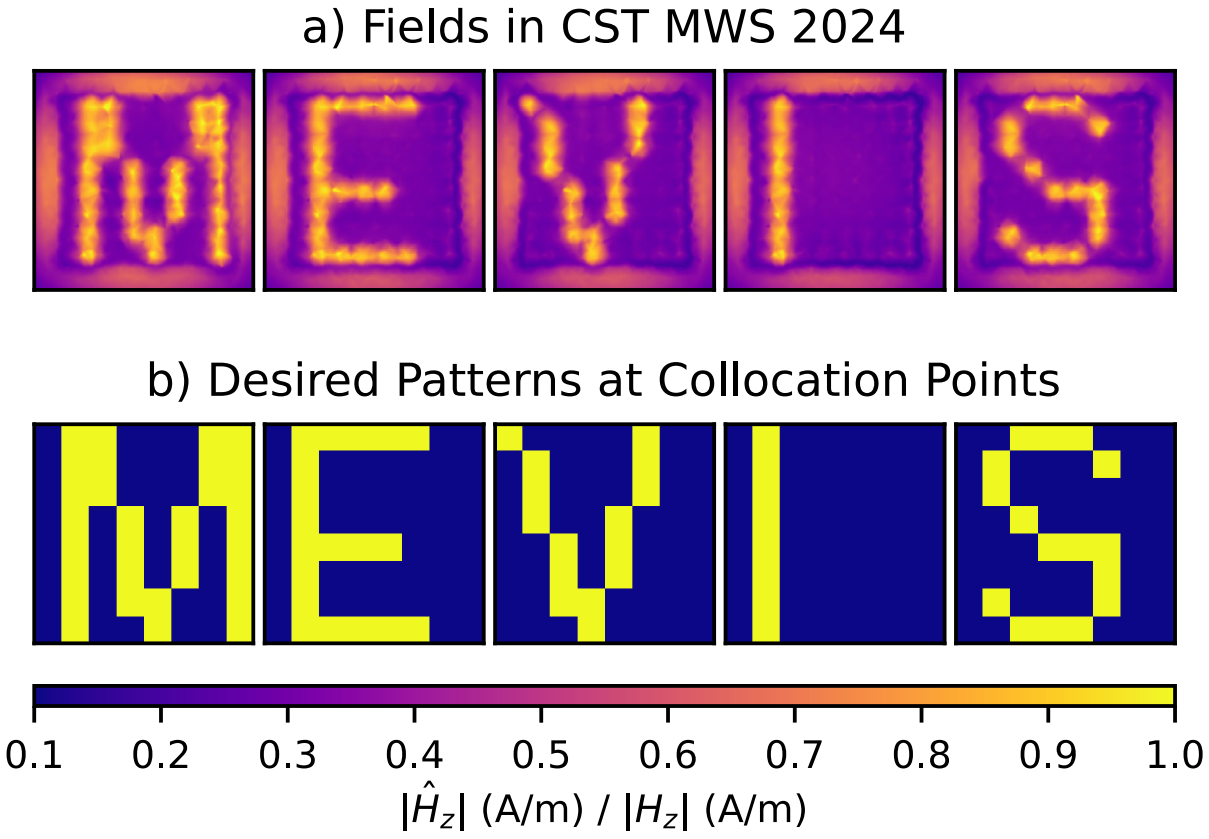

**FIGURE 10.** Demonstration of individual letters "M-E-V-I-S" realized on the 8 × 8 array. a) shows $|\hat{H}_z|$ as realized and verified in CST MWS relative to the target $|\mathbf{H}_z|$. A field-of-view of 300 × 300 mm is shown. The 8 × 8 array has an extent of 220 × 220 mm and is centered in the field of view. The plane was 0.8 times the diameter of the unit cell, which corresponds to 20 mm. Therefore, in this view, the edges of the array are clearly visible. b) shows $|\mathbf{H}_z|$ as target by the user in the collocation points.

(s. Fig. 10). Furthermore, the testing procedures verify and crosscheck the output of the differentiable simulator with the output of another non-differentiable cosimulator [18], thus guaranteeing agreement.

## V. DISCUSSION

The convergence of gradient descent depends on the choice of the initial parameters. In addition, given the inverse nature of the problem, one or more Hadamard conditions may be breached. Both are addressed in the following sections.

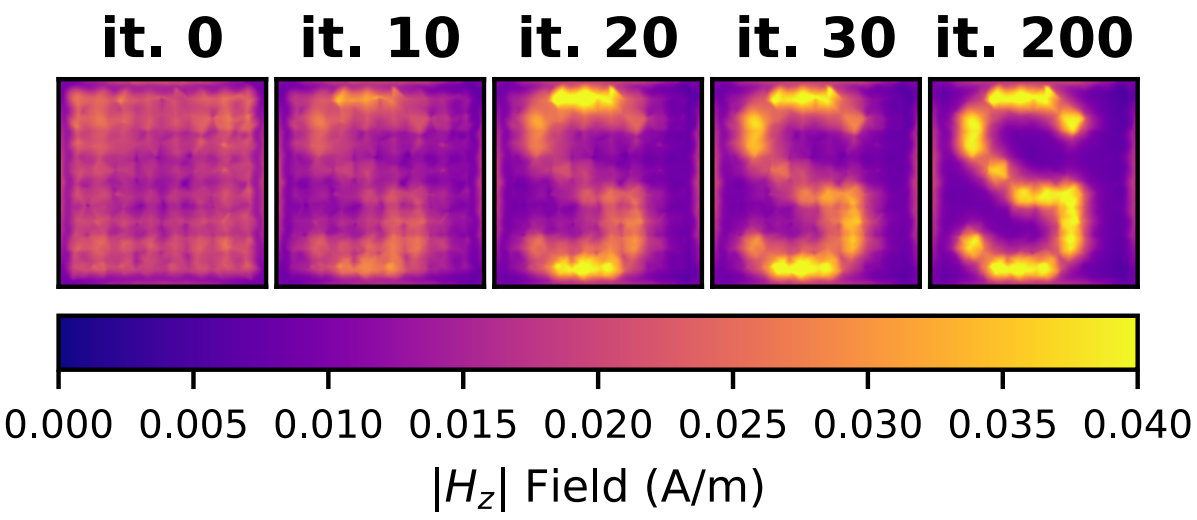

**FIGURE 11. Exemplary evolution of magnetic field distribution $|\mathbf{H_z}|$ of 8 × 8 array over iterations of gradient descent-based optimization procedure for letter "S".**

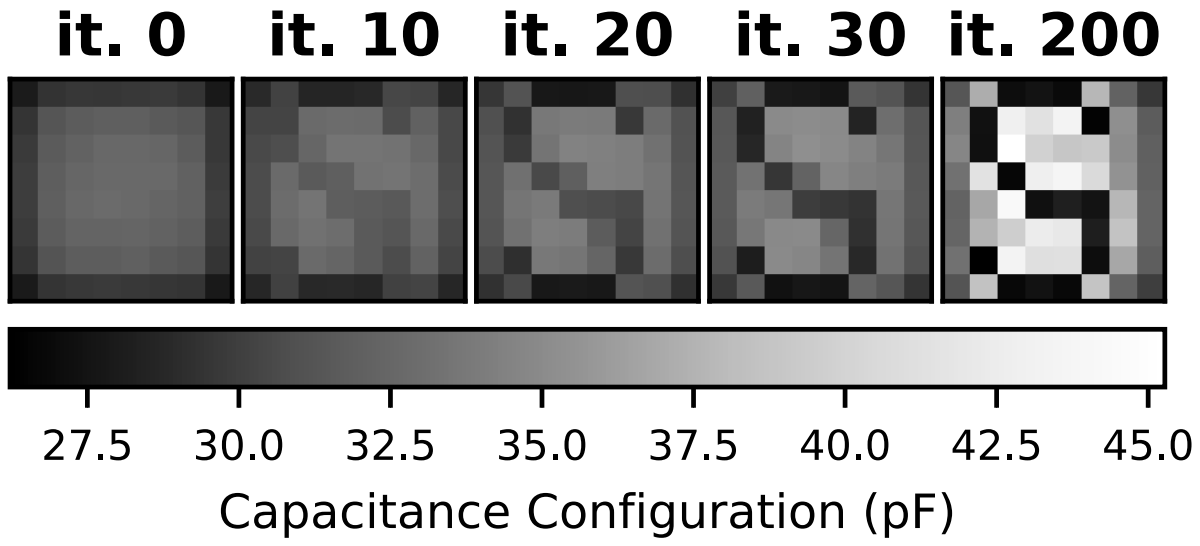

**FIGURE 12. Exemplary evolution of the capacitance configuration of an 8 × 8 array over iterations of gradient descent-based optimization procedure for letter "S".**

### A. RECONFIGURABLE UNIT-CELL

To assess whether $C_r$ is an effective starting point for the unit cell optimization, a parametric sweep is conducted on 100 individual initial parameters equidistantly sampled between 0.1 pF 30 pF. These individual initial parameters were then used to find solutions to 100 target $|\mathbf{H}_z|$ between 0.2 and 2.3 A/m. The resulting loss plot is shown in Fig. 13.

The loss plot in Fig. 13 exhibits discontinuities and can be separated into three different parts. $C_r = 8.39$ pF and $C_r = 12.13$ pF are distinctive points within the plot, marking the falling or rising edge of the loss. The initial parameter with the lowest overall MSE ($=1.26 \times 10^{-29}$) is approximately 10.96 pF. This represents significantly better convergence compared to, for example, an initial parameter of 12.42 pF, which produced an MSE of 1.39, despite the proximity of the two values. A visualization of the error over different target magnetic-field strengths is shown in Fig. 14.

While $C_r$ is in the range of parameters that lead to low overall errors, it does not lead to the lowest overall error within the given number of epochs.

Considering Fig. 15 helps in understanding how many solutions exist (if any) and what solutions can be reached from what starting point. At a glance, it is visible that when seeking any $|\mathbf{H}_z|$ below the global minimum and above the global maximum, the first Hadamard condition is violated as no solution exists within the given capacitance range. Within the global minimum and maximum of the loss plot in Fig. 13 multiple solutions potentially exist. Therefore, the second Hadamard condition is violated for some target field strengths. Consequently, the solution found depends on the initial parameters. The initial parameters can be varied systematically. If there are more than one low error solutions, then the second hadamard condition is infringed.

The solutions leading to the overall lowest error are in region II of the absolute and phase plot of $\mathbf{H}_z$ in Fig. 15.

- with respect to the phase plot one can observe that the best initial parameter can be found where the change of phase is the largest
- with respect to the absolute plot one can observe that the best initial parameter can be found where the range of capacitances can reach all $|\mathbf{H}_z|$.

Therefore, only within the range $8.39 \leq C \leq 12.13$ pF, a solution for all realizable $|\mathbf{H}_z|$ can be found. The reason for this lies in the behavior of gradient descent: vanilla gradient descent tries to decrease the error steadily and therefore converges towards a high error because, for example, for region III in Fig. 15 and Fig. 13 the error must increase before it decreases again. gradient descent is generally not able to go in the direction of higher errors for a larger number of epochs. Therefore, it cannot overcome areas of increasing errors and will settle for a local minimum.

The Adam optimizer can continue moving in a direction of increasing error for a limited time due to its momentum-like behavior. Therefore, overcoming shallow local minima is possible. This effect decayed if the uphill trend persisted. It may therefore not be possible to overcome steep local minima for prolonged number of epochs as it may be the case for regions I and III in Fig. 13. A combination of a local optimizer and a global optimizer may alleviate the constraints imposed by suboptimal initial parameters.

### B. HIGHER DIMENSIONAL MTS

A notable observation in the error distribution is the presence of a long-tailed distribution (see Fig. 7 f)), which could result from unsuitable parameter initialization or from the unphysical violation of the first Hadamard condition by the enhancement parameter $\beta$ in specific cases. Specifically, if $\beta$ breaches this condition, it will likely result in an increased error.

For systematic patterns, as shown in Fig. 8, at a high $\beta$, the gradient descent may attempt to converge towards a non-existent solution, leading to increased errors. Moreover, the observation point of each individual cell was positioned at 0.8 times their diameter. This allows nearby cells to influence the neighboring observation points. Maintaining a lower field enhancement in only one place, while concurrently sustaining a higher target field strength in many other places, could be harder to realize because of the influence of the other cells on the observation point. This may also apply for highly frequent spatially varying patterns. While this may seem like an adverse effect at first, this may allow constructive or destructive building of more continuous patterns in deeper layers.

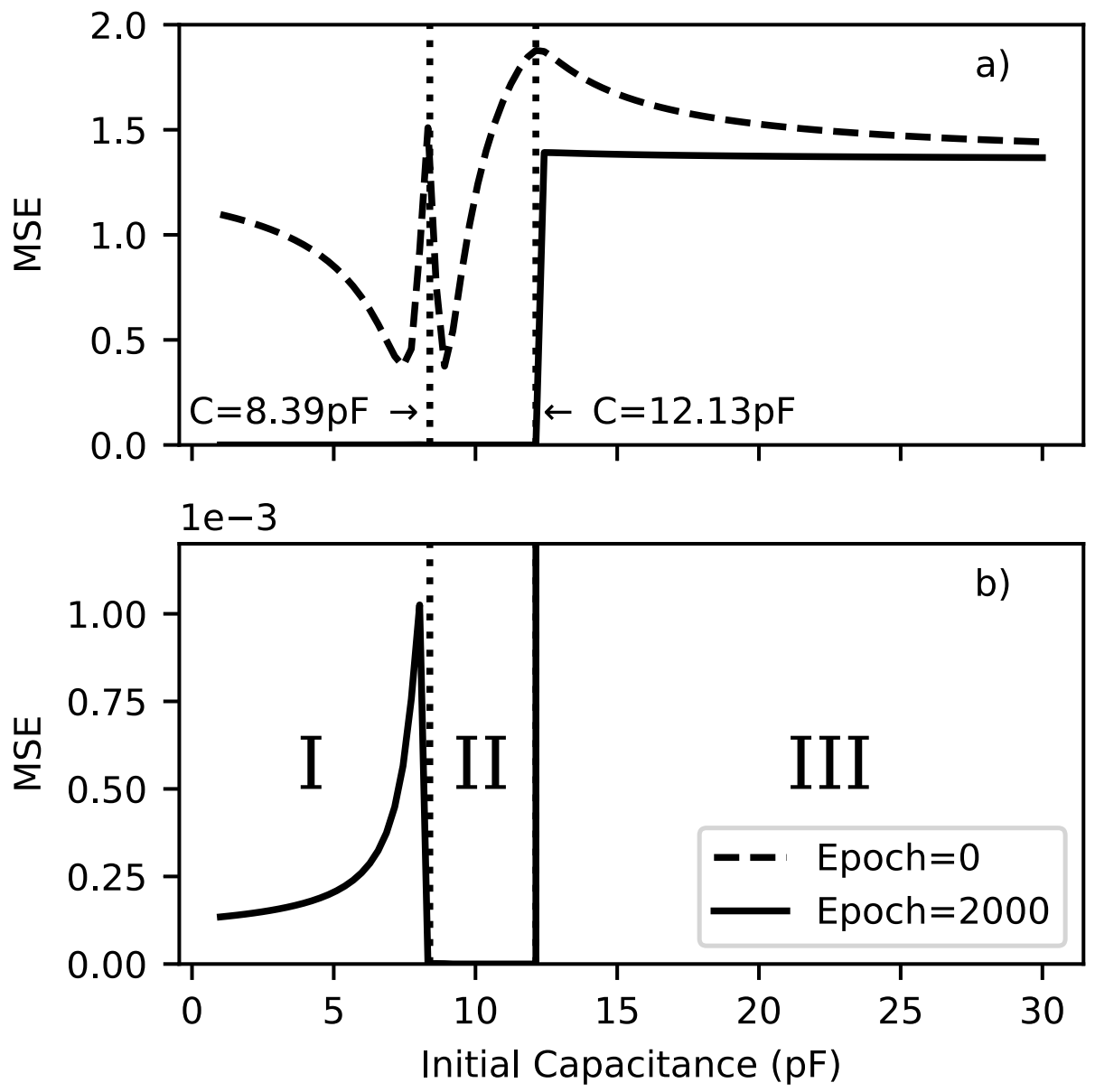

**FIGURE 13.** The initial capacitance configuration of the one unit cell setup is swept from 0.1 pF up to 30 pF and then used for optimization. The losses in the last epoch (epoch=2000) are visualized. The loss plot can be divided into three regions (I, II, and III). The capacitances in region II lead to the overall lowest error while capacitances in region III lead to the highest overall error.

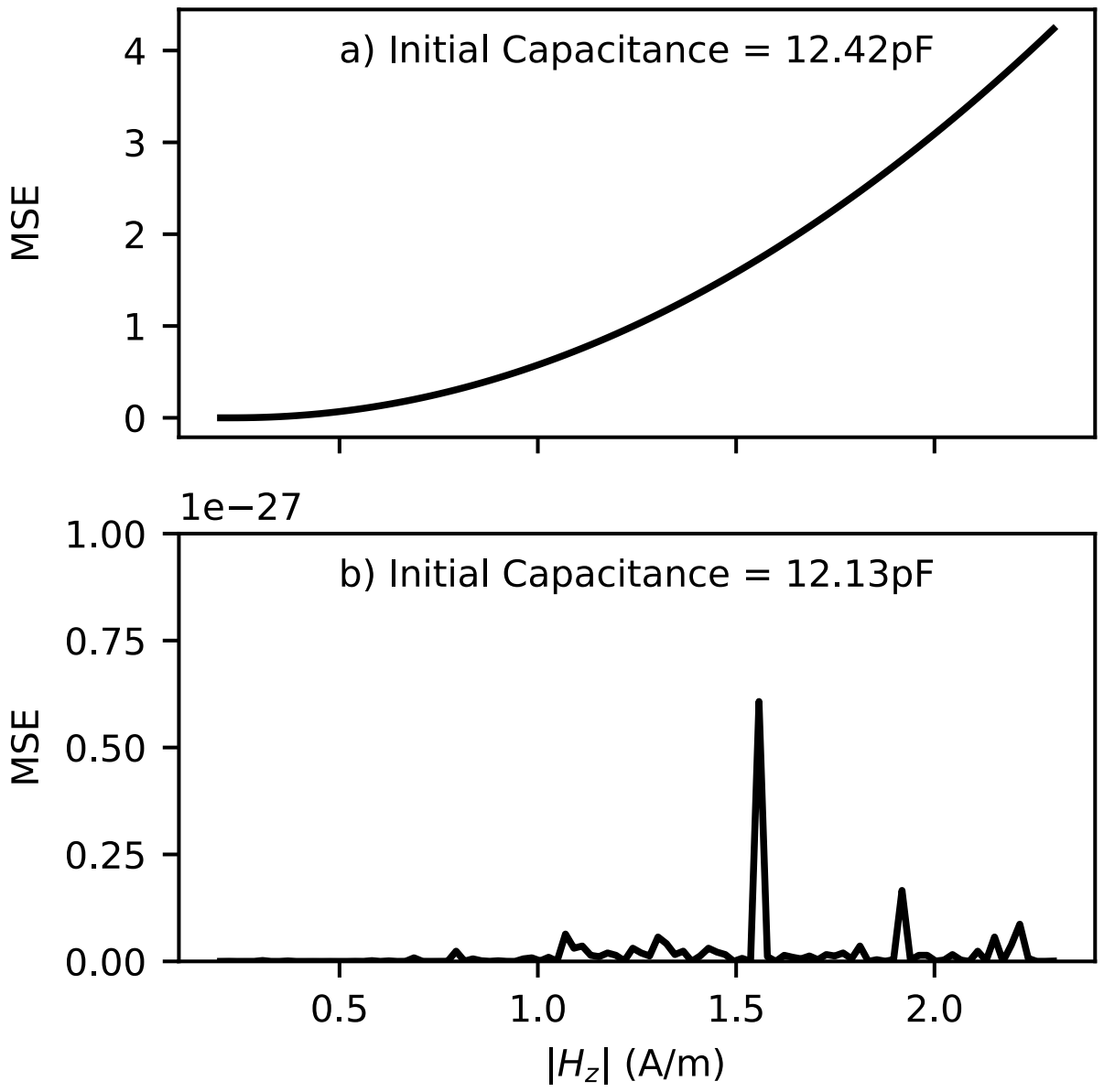

**FIGURE 14.** Difference in final error (epoch=2000) per sample between the initial capacitance set to 12.13 pF in a) which results in the highest overall error and initial capacitance set to 12.42 pF in b) which results in one of the lowest overall errors. For the majority of samples there is a more than $1 \times 10^{-27}$ difference in scale with respect to the errors.

The selection of the initial parameters critically affects the algorithm's convergent behavior towards an optimal solution. The initial parameter set chosen for a given field enhancement $\beta$ may not be suitable when applied to different configurations, thus hindering the effective optimization. Unfortunately, the discussion of initial parameters is not trivial in more dimensional cases, as the number of potential initial parameters scales exponentially with ${n_{Cs}}^{n_{cell}}$, where $n_{Cs}$ denotes the number of all possible capacitance states.

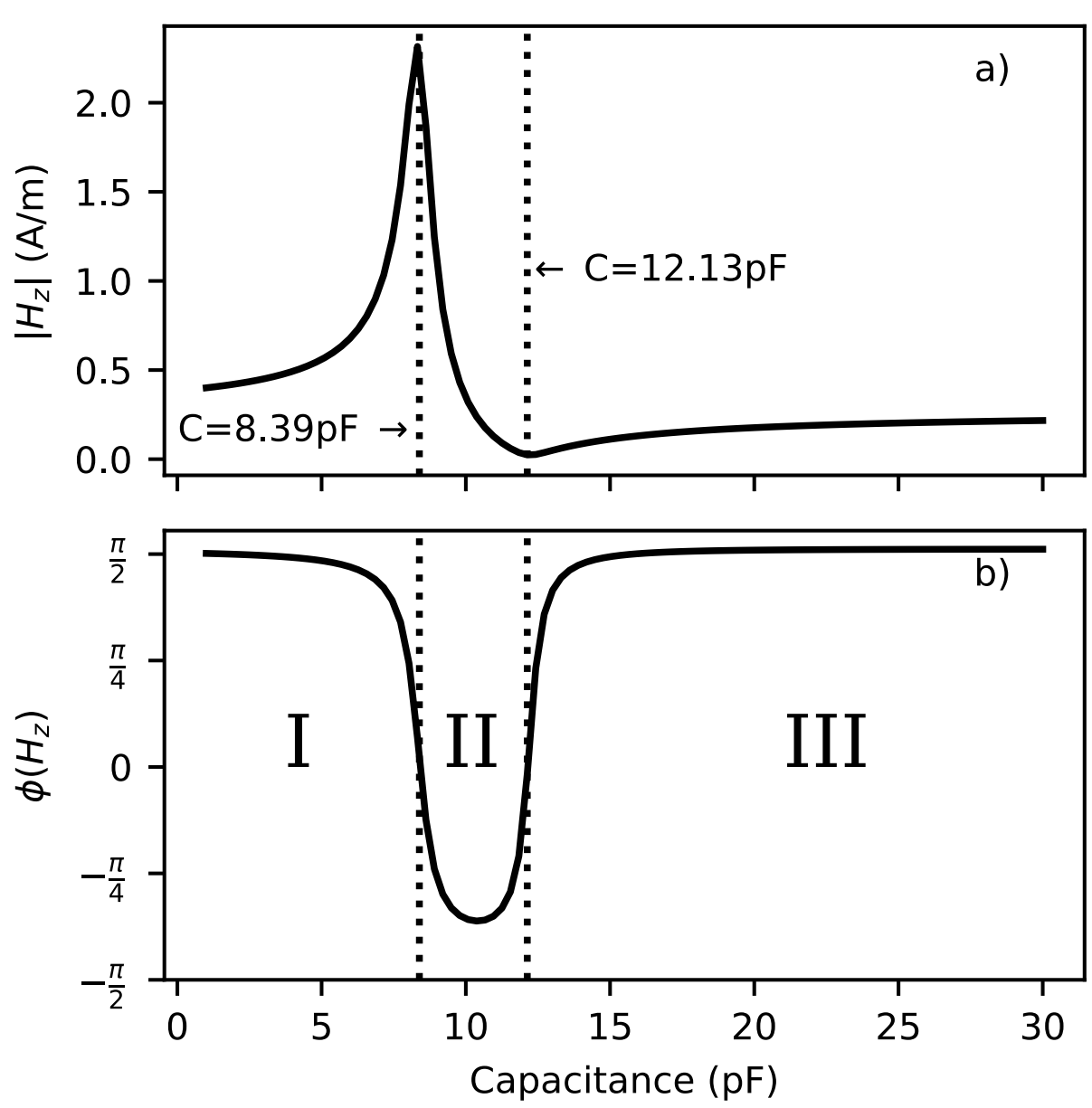

**FIGURE 15.** Variation of the capacitance and the effect on the a) absolute and b) phase of the magnetic field $|\mathbf{H_z}|$ for a single-cell setup. The plot illustrates the nonlinear relationship observed, highlighting the key points where significant changes in the magnetic field occur with respect to the capacitance. Only the capacitances in region II can realize all possible target magnetic field strengths. This correlates with low overall errors in Fig. 13.

Samples with greater spatial variation exhibited a higher error than those with less spatial variation. This supports the presumption that solutions that realize patterns with more spatial variation are harder or impossible to find, as they might not even exist.

Another issue becomes apparent when using the $8 \times 8$ structure: Any real number is allowed to be a possible solution. This allows parameter sets to contain unrealistically high or low capacitance values as a solution.

In the $8 \times 1$ configuration, the final capacitance values are closer to its initial parameters (spanning from 25.4 pF to 26.4 pF) with its lowest and highest value after optimization of 24.7 pF to 35.1 pF. Conversely, in the $8 \times 8$ case, gradient descent optimization trended towards much higher capacitance values, with a lowest value of 26.1 pF and a highest value of 61.2 pF. The initial parameters for the $8 \times 8$ configuration ranged from 27.8 pF to 32.9 pF. In this case, the increase in the mean and range of the initial configuration is noticeable. Therefore, the increased coupling leads to an inherent change in the behavior of the overall resonant structure. Letting the algorithm run 20000 epochs will lead to an even further decrease of the total error but also to an increase in some capacitance configurations (maximum C $\approx$ 450 pF). If optimization is performed with an existing structure in mind, a penalty function should be configured accordingly.

Learning rate scheduling may improve convergence speed depending on how far the initial solution is from a potential low error solution. If the initial solution is close to a potential low-error solution, a higher initial learning rate may result in worse intermediate results. Worse intermediate solutions may be worse than the initial solution and may make it impossible to find a good candidate solution. No gradient clipping was used as no numerical instability has been observed.

Bounded projection functions, such as the tanh projection, can limit the capacitance range during optimization. However, if the allowed range is too narrow, some realizable field distributions may become unattainable because the required capacitance values fall outside the bounds. The appropriate capacitance tuning range depends on the specific physical components selected by the user.

If the optimizer encounters boundary limits on one side of the range, it raises the question of whether better solutions may exist on the other side of the spectrum. This is analogous to moving from Region III to Region I in Fig. 13 and Fig. 15. It may be sufficient to set a large tuning range and simply clip values that exceed the physical limits, rather than strictly constraining the design space with a bounded projection.

In addition, optimization was conducted only on selected collocation points. Therefore, the patterns realized in Fig. 10 appear not to be smooth. Adding more collocation points may allow finer tuning for smoother patterns. Unfortunately, this also requires better knowledge of how to set additional collocation points, as they constrain one another. No low-error solution can be found if unrealistic field patterns are realized.

Although satisfactory solutions have been found for many of the scenarios considered, it must be emphasized that gradient descent techniques do not guarantee convergence to a global minimum. Instead, the algorithm may settle at local minima.

To avoid the pitfalls of local minima, particularly in high-dimensional parameter spaces for the initial parameters, more sophisticated optimization techniques, such as Bayesian optimization, as illustrated in [27], could be considered. Embedding the simulator in a deep learning pipeline is an additional option. The framework developed in this study can be seamlessly integrated with advanced techniques, allowing for the efficient embedding of the forward problem within a broader optimization landscape.

Given a suitable initial parameter, gradient descent can find the parameter configuration of a linear electromagnetic structure with exactly one tunable parameter that allows the realization of a target magnetic field strength $|\mathbf{H}_z|$. The solution depends on the initial parameters.

Despite these advancements, physical limitations remain, such as the potential deviation of simulations from real-world scenarios. One could establish a calibration routine that incorporates knowledge from real-world setups and accounts for differences to simulation.

The proposed approach supports the use of other gradient-based optimization methods and facilitates the embedding of the forward problem into other optimization procedures. For instance, the framework can serve as a foundation for further deep learning research, potentially being integrated into physics-informed deep learning pipelines that guide a neural network during training.

A resonant structure with 64 inductively coupled unit cells, each with one individually tunable parameter, can be optimized to achieve the target magnetic field strengths $|\mathbf{H}_z|$ at designated points. Specific letters [A-Z], digits [0-9], homogenization, and one Hadamard pattern can be realized.

It was successfully demonstrated that a resonant structure with eight inductively coupled unit cells, each with individually tunable parameters, can be optimized to achieve target magnetic field enhancements $\beta$ at designated points. The structure could realize arbitrary enhancement patterns, Hadamard patterns, homogenization, as well as peaks and troughs at one-cell length, within a reasonable range. These simple examples are intended to demonstrate the newly gained possibility of using gradient descent in the optimization of reconfigurable electromagnetic structures.

## VI. CONCLUSION

A CUDA-enabled PyTorch-based framework for the accurate and efficient calculation of electric and magnetic field distribution with respect to tunable electrical parameters has been presented and made publicly available on github.com. Leveraging automatic differentiation, deliberate and direct control of the magnetic field's magnitude (specifically the component orthogonal to the resonant structure) was demonstrated.

It is found that ...

- it is possible to precisely control the magnitude of the magnetic field normal to the center of each individual unit cell.
- the successful realization of target homogenized field strengths and specific field distributions (such as peaks) at designated spatial points with low error is feasible when considering discussed constraints.

It is therefore possible to infer capacitance configurations that realize a target magnetic field distribution at given points in space. Examples indicating the fulfillment of the study's objective are shown. These findings imply that a suitable configuration for a reconfigurable MTS in MRI can be determined in order to optimize for a specific imaging goal.

The suggested approach paves the way for advanced electromagnetic control strategies and provides an extendable platform for further exploration and innovation in this field.

## DECLARATION OF INTERESTS

The authors declare that they have no known competing financial interests or personal relationships that could have influenced the work reported in this paper.

## DECLARATION OF GENERATIVE AI AND AI-ASSISTED TECHNOLOGIES IN THE WRITING PROCESS

During the preparation of this manuscript, the authors partially used OpenAI's ChatGPT4o service to help formulate individual sentences. Github Copilot was partially used for code revision and comment generation. After using these services, the authors reviewed and edited the content as needed and take full responsibility for the content of the published article.

## ACKNOWLEDGMENT

The authors would like to thank Umberto Zanovello, whose work on CoSimPy [18] inspired their research and the development of this simulator.

## REFERENCES

[1] A. Shchelokova, R. Schmidt, A. Slobozhanyuk, T. Kallos, A. Webb, and P. A. Belov, "Enhancement of magnetic resonance imaging with metasurfaces: From concept to human trials," in *Proc. 11th Int. Congr. Eng. Mater. Platforms Novel Wave Phenomena (Metamaterials)*, Aug. 2017, pp. 31–33.

[2] M. J. Freire, R. Marques, and L. Jelinek, "Experimental demonstration of a $\mu$=-1 metamaterial lens for magnetic resonance imaging," *Appl. Phys. Lett.*, vol. 93, no. 23, Dec. 2008, Art. no. 231108. [Online]. Available: https://aip.scitation.org/doi/full/10.1063/1.3043725

[3] M. J. Freire, L. Jelinek, R. Marques, and M. Lapine, "On the applications of metamaterial lenses for magnetic resonance imaging," *J. Magn. Reson.*, vol. 203, no. 1, pp. 81–90, Mar. 2010. [Online]. Available: https://www.sciencedirect.com/science/article/pii/S1090780709003607

[4] J. M. Algarín, M. J. Freire, F. Breuer, and V. C. Behr, "Metamaterial magnetoinductive lens performance as a function of field strength," *J. Magn. Reson.*, vol. 247, pp. 9–14, Oct. 2014.

[5] E. A. Brui, A. V. Shchelokova, M. Zubkov, I. V. Melchakova, S. B. Glybovski, and A. P. Slobozhanyuk, "Adjustable subwavelength metasurface-inspired resonator for magnetic resonance imaging," *Phys. Status Solidi A*, vol. 215, no. 5, Mar. 2018, Art. no. 1700788. [Online]. Available: https://onlinelibrary.wiley.com/doi/abs/10.1002/pssa.201700788

[6] A. V. Shchelokova, A. P. Slobozhanyuk, I. V. Melchakova, S. B. Glybovski, A. G. Webb, Y. S. Kivshar, and P. A. Belov, "Locally enhanced image quality with tunable hybrid metasurfaces," *Phys. Rev. Appl.*, vol. 9, no. 1, Jan. 2018, Art. no. 014020.

[7] E. Stoja, S. Konstandin, D. Philipp, R. N. Wilke, D. Betancourt, T. Bertuch, J. Jenne, R. Umathum, and M. Günther, "Improving magnetic resonance imaging with smart and thin metasurfaces," *Sci. Rep.*, vol. 11, no. 1, p. 16179, Aug. 2021. [Online]. Available: https://www.nature.com/articles/s41598-021-95420-w

[8] T. H. Hand and S. A. Cummer, "Controllable magnetic metamaterial using digitally addressable split-ring resonators," *IEEE Antennas Wireless Propag. Lett.*, vol. 8, pp. 262–265, 2009.

[9] G. Oliveri, D. H. Werner, and A. Massa, "Reconfigurable electromagnetics through metamaterials—A review," *Proc. IEEE*, vol. 103, no. 7, pp. 1034–1056, Jul. 2015.

[10] Q. He, S. Sun, and L. Zhou, "Tunable/reconfigurable metasurfaces: Physics and applications," *Research*, vol. 2019, Jan. 2019, Art. no. 1849272, doi: 10.34133/2019/1849272.

[11] H. Wang, H.-K. Huang, Y.-S. Chen, and Y. Zhao, "On-demand field shaping for enhanced magnetic resonance imaging using an ultrathin reconfigurable metasurface," *View*, vol. 2, no. 3, Feb. 2021, Art. no. 20200099.

[12] M. Lippke, E. Stoja, D. Philipp, S. Konstandin, J. Jenne, T. Bertuch, and M. Gunther, "Investigation of a digitally-reconfigurable metasurface for magnetic resonance imaging," in *Proc. 52nd Eur. Microw. Conf. (EuMC)*, Sep. 2022, pp. 668–671.

[13] E. Stoja, D. Philipp, S. Konstandin, J. Jenne, T. Bertuch, and M. Günther, "Reconfigurable metasurfaces and new imaging paradigms in magnetic resonance imaging," in *Proc. 17th Eur. Conf. Antennas Propag. (EuCAP)*, Mar. 2023, pp. 1–4.

[14] J. Müller, M. Falchi, E. Stoja, S. Konstandiri, M. Gunther, D. Brizi, P. Usai, A. Monorchio, and D. Philipp, "Deep-learning optimized reconfigurable metasurface for magnetic resonance imaging," in *Proc. 18th Eur. Conf. Antennas Propag. (EuCAP)*, Mar. 2024, pp. 1–5.

[15] M. Eisterer, "The significance of solutions of the inverse biot–savart problem in thick superconductors," *Superconductor Sci. Technol.*, vol. 18, no. 2, pp. S58–S62, Dec. 2004.

[16] R. Teyber, L. Brouwer, J. Qiang, and S. Prestemon, "Inverse biot–savart optimization for superconducting accelerator magnets," *IEEE Trans. Magn.*, vol. 57, no. 9, pp. 1–7, Sep. 2021.

[17] A. Beqiri, J. W. Hand, J. V. Hajnal, and S. J. Malik, "Comparison between simulated decoupling regimes for specific absorption rate prediction in parallel transmit MRI," *Magn. Reson. Med.*, vol. 74, no. 5, pp. 1423–1434, Nov. 2015.

[18] U. Zanovello, F. Seifert, O. Bottauscio, L. Winter, L. Zilberti, and B. Ittermann, "CoSimPy: An open-source Python library for MRI radiofrequency coil EM/circuit cosimulation," *Comput. Methods Programs Biomed.*, vol. 216, Apr. 2022, Art. no. 106684.

[19] D. Brizi and A. Monorchio, "An analytical approach for the arbitrary control of magnetic metasurfaces frequency response," *IEEE Antennas Wireless Propag. Lett.*, vol. 20, pp. 1003–1007, 2021.

[20] R. Newbury, J. Collins, K. He, J. Pan, I. Posner, D. Howard, and A. Cosgun, "A review of differentiable simulators," *IEEE Access*, vol. 12, pp. 97581–97604, 2024. [Online]. Available: https://ieeexplore.ieee.org/document/10589638

[21] M. Kozlov and R. Turner, "Fast MRI coil analysis based on 3-D electromagnetic and RF circuit co-simulation," *J. Magn. Reson.*, vol. 200, no. 1, pp. 147–152, Sep. 2009.

[22] CST AG. *CST Studio Suite 2024 Help*. Accessed: Jul. 3, 2024. [Online]. Available: https://www.3ds.com/products-services/simulia/products/cst-studio-suite/

[23] K. B. Petersen and M. S. Pedersen. (2012). *The Matrix Cookbook*. Accessed: Sep. 27, 2025. [Online]. Available: https://www.math.uwaterloo.ca/~hwolkowi/matrixcookbook.pdf

[24] S. M. Selby, *Standard Mathematical Tables*. Boca Raton, FL, USA: CRC Press, 1974.

[25] ZMT Zurich MedTech AG. *Sim4Life*. Accessed: Jul. 3, 2024. [Online]. Available: http://www.zmt.swiss/sim4life/

[26] D. P. Kingma and J. Ba, "Adam: A method for stochastic optimization," 2014, *arXiv:1412.6980*.

[27] R. Antonova, J. Yang, K. M. Jatavallabhula, and J. Bohg, "Rethinking optimization with differentiable simulation from a global perspective," 2022, *arXiv:2207.00167*.

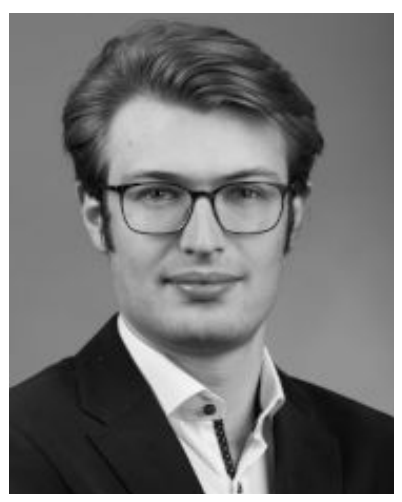

**JOHANNES MÜLLER** (Member, IEEE) was born in Lower Saxony, Germany, in 1996. He received the Bachelor of Science and Master of Science degrees in electrical engineering and management from the University of Bremen, Germany, in 2019 and 2022, respectively.

He is a Scientific Researcher. Throughout his academic journey, he gained extensive practical experience in scientific projects and industry. For finance applications, he applied machine learning techniques to condensed time-series data. In the field of communication engineering, he applied an approach to converge Least Squares (LS) to Linear Minimum mean Square Error (LMMSE) solutions using multi-layered neural networks. For his master's thesis, he focused on the application of Optical Coherence Tomography (OCT) in the industry, deriving topographical features of weld seam characteristics for practical use. His research interests include machine learning, deep-learning-optimized systems, complex systems, electromagnetism, communication and information engineering, and application-oriented research.

Mr. Müller is a member of the VDE (Verband der Elektrotechnik Elektronik Informationstechnik) and IEEE professional society.

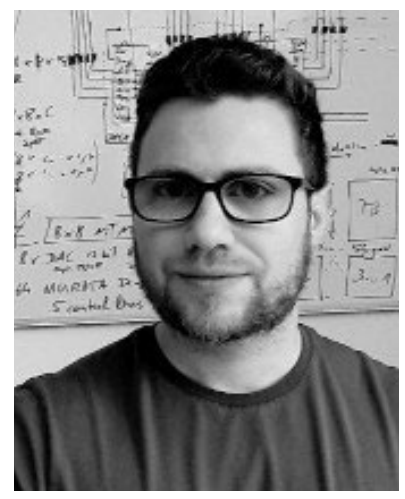

**DENNIS PHILIPP** received the B.Sc. degree in physics from the University of Bremen, in 2012, the M.Sc. degree in physics from the Carl von Ossietzky University of Oldenburg, Germany, in 2014, and the doctoral degree in theoretical gravitational physics from the University of Bremen, Germany, in 2018.

From 2014 to 2018, he was a Research Fellow and the Doctoral Candidate with ZARM, focusing on the theoretical aspects of general relativity. He is currently a Postdoctoral Researcher with the Center of Applied Space Technology and Microgravity (ZARM) and Fraunhofer MEVIS, Bremen. He is a Scientific Researcher. He has held tutor positions in mathematics and physics with the Universities of Bremen and Oldenburg. His research interests include quantum metrology, gravitational field modeling, clock effects in spacetime, MRI, and metamaterials.

Dr. Philipp is a member of the ISMRM, ESMRM, EGU, and DPG. He serves as a reviewer for journals, such as Classical and Quantum Gravity, European Journal of Physics, and Nature Scientific Reports. He was awarded the Bremen Study Prize, in 2020 for his dissertation.

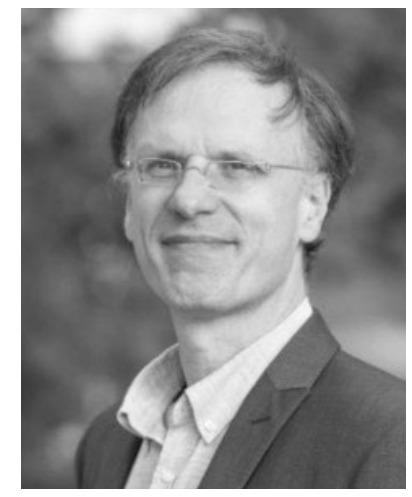

**MATTHIAS GÜNTHER** received the Dr. rer. nat. and Diploma degrees in physics from Heidelberg University, and the Prediploma degree in economic sciences from FernUni Hagen.

He is currently a Professor of MR-Physics with the University of Bremen and the Head of MR-Imaging with Fraunhofer MEVIS, Bremen, where he also serves on the Management Board. He is a Scientific Researcher. His prior roles include a Group Leader with the Neurological Clinic, University Clinic Mannheim, a Senior Scientist with Advanced MRI Technologies, Sebastopol, CA, a Postdoctoral Researcher with the Deutsches Krebsforschungszentrum, Heidelberg, and a Guest Scientist with UC Berkeley. During his Diploma degree, he received the postgraduate certification in medical physics. His research interests include quantitative MR-perfusion imaging, arterial spin labeling, and dynamic sequence development for clinical applications.

Prof. Günther is a member of the International Society for Magnetic Resonance in Medicine and European Society for Magnetic Resonance in Medicine and Biology. He serves on the Editorial Board for IEEE Transactions on Medical Imaging, and reviews for European Research Council, DFG, and SNSF. He has received multiple research fellowships and contributes to Fraunhofer and IEEE technical committees.